\documentclass[useAMS,usenatbib]{mn2e} 
\usepackage{times}

\usepackage{subfigure}
\usepackage{rotating,caption}
\usepackage{lscape}
\usepackage{amsmath}
\usepackage{amssymb}
\usepackage{natbib}
\usepackage{rotating}
\usepackage[bookmarks=true]{hyperref}
\usepackage{color}
\definecolor{darkgreen}{rgb}{0,0.5,0}
\definecolor{darkblue}{rgb}{0,0,0.55}
\hypersetup{colorlinks,linkcolor=blue,filecolor=darkgreen,citecolor=blue}

\newcommand{\msun}{$~M_{\odot}$}

\newcommand{\HII}{H~{\sc ii}}

\begin{document}

\title[IRAS\,16148-5011]
      {Study of morphology and stellar content of the Galactic \HII\, region IRAS\,16148-5011}  

\author[K. K. Mallick et al.] 
       {K. K. Mallick,$^1$\thanks{E-mail: kshitiz@tifr.res.in} D. K. Ojha,$^1$ M. Tamura,$^2$ H. Linz,$^3$, M. R. Samal,$^4$ 
        \newauthor 
        and S. K. Ghosh$^{1,5}$ \\
       $^1$ Department of Astronomy and Astrophysics, Tata Institute of Fundamental Research, Homi Bhabha Road, Colaba, \\
            Mumbai 400 005, India \\
       $^2$ National Astronomical Observatory of Japan, Mitaka, Tokyo 181-8588, Japan \\
       $^3$ Max Planck Institute for Astronomy, K{\"o}nigstuhl 17, D-69117 Heidelberg, Germany \\
       $^4$ Aix Marseille Universit\'e, CNRS, LAM (Laboratoire d'Astrophysique de Marseille) UMR 7326, 13388 Marseille, France \\
       $^5$ National Centre for Radio Astrophysics, Tata Institute of Fundamental Research, Pune 411 007, India \\ 
       }

\date{  }

\maketitle

\begin{abstract}
An investigation of the IRAS\,16148-5011 region - a cluster at a distance of 3.6\,kpc - is presented here, carried out using 
multiwavelength data in near-infrared (NIR) from the 1.4\,m Infrared Survey Facility telescope, mid-infrared (MIR) from 
the archival \textit{Spitzer} GLIMPSE survey, far-infrared (FIR) from the \textit{Herschel} archive, and low-frequency radio 
continuum observations at 1280 and 843\,MHz from the Giant Metrewave Radio Telescope (GMRT) and Molonglo Survey archive, 
respectively. A combination of NIR and MIR data is used to identify 7 Class\,I and 133 Class\,II sources in the region. 
Spectral Energy Distribution (SED) analysis of selected sources reveals a 9.6\,\msun\, high-mass source embedded in nebulosity. 
However, Lyman continuum luminosity calculation using radio emission - which shows a compact \HII\, region - indicates the 
spectral type of the ionizing source to be earlier than B0-O9.5. 
Free-free emission SED modelling yields the electron density as 138\,cm$^{-3}$, and thus the mass of the ionized hydrogen as 
$\sim$\,16.4\,\msun. 
Thermal dust emission modelling, using the FIR data from \textit{Herschel} and performing modified blackbody fits, helped us 
construct the temperature and column density maps of the region, which show peak values of 30\,K and 
3.3$\times$10$^{22}$\,cm$^{-2}$, respectively. The column density maps reveal an A$_V >$\,20\,mag extinction associated with the 
nebular emission, and weak filamentary structures connecting dense clumps. The clump associated with this IRAS object is found to 
have dimensions of $\sim$\,1.1\,pc$\times$0.8\,pc, and a mass of 1023\,\msun. 
\end{abstract}


\section{Introduction}  
\label{section_Introduction} 

Most star formation activity is known to take place in clusters \citep{lada03}, and as such, observational studies 
of young embedded stellar cluster regions are imperative, because they serve as a template to further investigate various 
associated processes and their signatures. Due to the youth of such regions and the fact that their natal medium has still not 
been dispersed, these clusters can be used to scrutinize various theories related to star formation, stellar cluster dynamics, 
as well as stellar and cloud evolution. Of even more importance are the cluster regions which harbour high-mass stars, partly 
because such regions are few and far between, and more so as high-mass star formation is not very well understood 
\citep{zinnecker07}. 

\begin{figure*}
\centering
\subfigure
{
\includegraphics[scale=0.4]{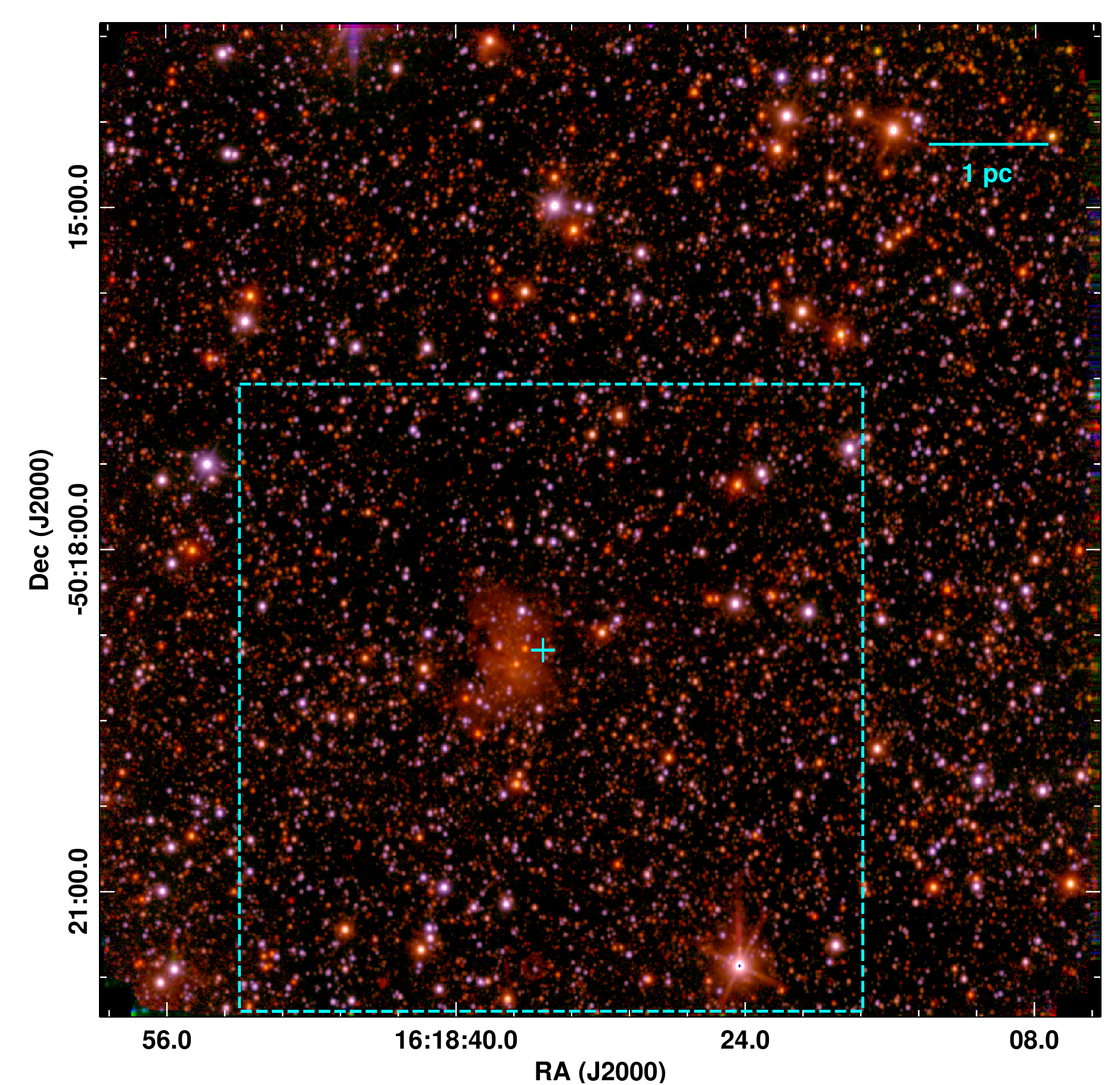}
}
\caption{Colour composite image of the IRAS\,16148-5011 region, made using the $K_s$ (red), $H$ (green), and $J$ (blue) band 
images. The area of analysis in this paper has been marked by the dashed rectangle. Plus symbol marks the IRAS catalogue position 
of IRAS\,16148-5011. The scale bar shows 1\,pc extent at a distance of 3.6\,kpc.}  
\label{fig_Introduction_ColourComposite} 
\end{figure*}

IRAS\,16148-5011 is an infrared nebula (Fig. \ref{fig_Introduction_ColourComposite}) in the southern sky 
($\alpha_{2000} = 16^h18^m35.2^s$, $\delta_{2000} = -50^{o}18\arcmin53\arcsec$) associated with which is an IR cluster found 
using the Two Micron All-Sky Survey (2MASS) data by \citet{dutra03}. 
It is located at the Galactic plane ($l \sim 333.047^{o}, b \sim +0.037^{o}$) and is in the vicinity of the well-known 
star-forming region RCW\,106. Though other star-forming regions are present nearby, IRAS\,16148-5011 appeared to be a relatively 
isolated region in past mappings \citep{karnik01,mookerjea04}. 
Kinematic distance estimates to this region vary from $\sim$ 3.3-11.9\,kpc \citep[near- and far-distance estimates;][]{molinari08}, 
and we adopt the distance of 3.6\,kpc from \citet[][]{lumsden13} (based on the spectrophotometric distance of a prominent 
source - \textquotedblleft G333.0494$+$00.0324B\textquotedblright\, in their nomenclature - associated with the central 
nebula\footnote{see http://rms.leeds.ac.uk/}) for our work. 
The compilation of \citet{lumsden13} also reveals other nearby regions at this distance, providing preliminary indications 
that this could be a part of larger complex.
In the early analyses of \citet{haynes79} \citep[also see][]{chan96}, radio continuum emission at 6\,cm was detected, with peak in 
the neighbourhood 
($\gtrsim$\,1\arcmin\,, positional accuracy $\sim$\,30\arcsec\,) of this region, and it was predicted to 
be harbouring massive young stellar objects. The IRAS colour analysis by \citet{macleod98} was also found to be consistent with 
that for an \HII\, region, and \citet{molinari08} - using the IRAS \textquotedblleft [25-12]\textquotedblright\, colour value - 
have suggested that this region is among the younger IRAS detected regions. 
A high-mass stellar source was detected by \citet{grave09} with the help of spectral energy distribution (SED) fitting 
using NIR to millimeter data. Analyses of this region at 1.2\,mm dust continuum emission and at molecular lines have revealed the 
presence of dense gas, with a large column density, as well as massive clumps. The total luminosity 
estimate ($\sim$\,4.4$\times$10$^4$\,L$_{\odot}$, by integrating IRAS flux densities) has also been found to be well in the regime
of high-mass stellar objects \citep{beltran06, fontani05}. 
Therefore, taking into account these characteristics, this region makes a good candidate to carry out an investigation 
for understanding the morphology and stellar population. However, since this is a possible \HII\, region with 
an embedded cluster, multiwavelength 
observations are required to fully discern this region's various constituents and how they relate to each other. In this paper, 
we have tried to accomplish this using deep NIR observations, archival MIR \textit{Spitzer} data, archival FIR \textit{Herschel} 
data, and low-frequency radio continuum observations.  

In Section \ref{section_ObsAndDataReduction}, we detail the various observations and the corresponding data analysis procedures, 
followed by an examination of the stellar population in the region in Section \ref{section_StellarPopulation}. The morphology 
of the region is discussed in Section \ref{section_Morphology}. Discussion and conclusions are presented in 
Sections \ref{section_Discussion} and \ref{section_conclusion}, respectively.

\section{Observations and data reduction} 
\label{section_ObsAndDataReduction}

\subsection{Near-Infrared Observations}
\label{section_NIR_Photometry}

NIR photometric observations in $J$\,(1.25\,$\mu$m), $H$\,(1.63\,$\mu$m), and $K_s$\,(2.14\,$\mu$m) 
bands (centered on $\alpha_{2000} \sim 16^h18^m31^s$, $\delta_{2000} \sim -50^{o}17^{'}32^{''}$) were carried out on 
2004 July 29 using the 1.4\,m Infrared Survey Facility (IRSF) telescope, South Africa. 
The observations were taken with the help of the Simultaneous InfraRed Imager for Unbiased Survey (SIRIUS) instrument, 
a three colour simultaneous camera mounted at the f/10 Cassegrain focus of the telescope. SIRIUS is equipped with 
three 1k$\times$1k HgCdTe arrays, each of which, with a pixel scale of 0.45\arcsec\,, provide a field of view (FoV) 
of 7.8\arcmin\,$\times$7.8\arcmin\,. Further details can be obtained from \citet{nagashima99} and \citet{nagayama03}. 
Five sets of frames, with each set containing observations at ten dithered positions (exposure time of 5\,s at each dither 
position), were obtained (i.e. total exposure time = 5$\times$10$\times$5\,s = 250\,s, in each band). The sky conditions 
were photometric, with a seeing size of $\sim$ 1.35\arcsec\,. In addition to the target 
field, a sky region ($\sim$\,10\arcmin\, to the north of target region) and the standard star P9172 \citep{persson98} were 
also observed. 

Following a standard data reduction procedure, which involved bad pixel masking, dark subtraction, flat-field correction, 
sky subtraction, combining dithered frames, and astrometric calibration, point spread function (psf) photometry 
was carried out using the {\sc allstar} algorithm of the {\sc daophot} package in {\sc iraf}. About 11-13 sources were 
used to construct the psf for each band. Finally the instrumental magnitudes were calibrated 
using the standard star P9172. The astrometric calibration rms obtained was $<$ 0.05\arcsec, and the median photometric 
error $<$ 0.05\,mag. On comparing our catalogues with the 2MASS catalogues, 
sources with $K_s$ magnitude $\leq$ 9.5 were found to be saturated, and hence had their $J, H,$ and $K_s$ magnitudes 
replaced by the corresponding 2MASS magnitudes. The final NIR catalogue is used to identify the young stellar objects 
(YSOs) in the region. The area of study in this paper is about 5.5\arcmin$\times$5.5\arcmin\, encompassing the nebular cloud
region, centered on $\alpha_{2000} \sim 16^h18^m35^s$, $\delta_{2000} \sim -50^{o}19^{'}18^{''}$.

Completeness limits were calculated for all three bands by carrying out artificial star experiments using the {\sc addstar}
package in {\sc iraf}. A fixed number of stars were added in each 0.5 magnitude bin and analysis carried out to see how 
many stars are detected. The ratio of the number of detected stars to the number of added stars gives us the completeness 
fraction as a function of magnitude. The 90\% completeness limit was thus calculated as 16.6, 15.8, and 15.5 for the 
$J, H,$ and $K_s$ bands, respectively.

\subsection{Radio Continuum Observations} 
\label{section_RadioObs} 

Radio continuum observations at 1280\,MHz were obtained on 2012\,November\,09 using the Giant Metrewave Radio Telescope 
(GMRT) array. The GMRT array consists of 30 antennae arranged in an approximate Y-shaped configuration, with each antenna 
having a diameter of 45\,m. This translates to a primary beam-size of 26.2\arcmin\, at 1280\,MHz. 
A central region of $\sim$1\,km$\times$1\,km contains 12 randomly distributed antennae, while the remaining 18 are 
along the 
three radial arms (6 along each arm) which extend upto $\sim$14\,km. Details about the GMRT can be found in \citet{swarup91}. 

For our observations, the Very Large Array (VLA) phase and flux calibrators \textquoteleft 1626-298\textquoteright\, and 
\textquoteleft 3C286\textquoteright\,, respectively, were used. The total observation time (including the calibrators) was about 
3.5\,hrs, limited by the low declination of the source. Data reduction was carried out using the {\sc aips} software. Initial   
steps involved flagging the bad data (carried out using a combination of \textquoteleft{\sc vplot-uvflg}\textquoteright\, and 
\textquoteleft{\sc tvflg}\textquoteright\, tasks) and calibration (carried out using 
\textquoteleft{\sc calib-getjy-clcal}\textquoteright\,). After a few iterations of flagging and calibration, the source data 
was \textquoteleft{\sc split}\textquoteright\, from the whole, and was used for imaging using the task 
\textquoteleft{\sc imagr}\textquoteright\,. A few rounds of (phase) self-calibration were also carried out using the task 
\textquoteleft{\sc calib}\textquoteright\, to remove any ionospheric phase distortion effects. To check that the flux 
calibration was done correctly, the image of the flux calibrator was constructed, and its flux determined and checked against 
literature values. 
The final (target) source images were rescaled to take into account the system temperature corrections for the GMRT. These 
corrections 
are required as, at the Galactic plane, the large amount of radiation at meter wavelengths increases the effective temperature 
of the antennae. This was done as follows. Using the sky temperature map of \citet{haslam82} at 408\,MHz, and the spectral 
index of -2.6 given therein, the temperature towards this region was obtained, 
i.e. $T_{frequency}=T_{408} \times (frequency/408\,{\rm MHz})^{-2.6}$, where $T_{408}$ is the temperature at 408\,MHz and 
$T_{frequency}$ is the temperature at required $frequency$ (1280\,MHz here). The images 
were subsequently rescaled by a factor given by $(T_{frequency} + T_{sys})/T_{sys}$, i.e. the ratio of temperature towards 
the target to that towards the flux calibrator. $T_{sys}$, the system temperature, was obtained from the GMRT 
manual\footnote{http://gmrt.ncra.tifr.res.in/gmrt\_hpage/Users/Help/help.html}. 

In addition to the GMRT observations, we also obtained the archival first epoch Molongolo Galactic Plane Survey (MGPS) data 
at 843\,MHz\footnote{http://www.physics.usyd.edu.au/sifa/Main/MGPS1} \citep{green99}. The MGPS observations were carried out 
using the Molonglo Observatory Synthesis Telescope (MOST)\footnote{The MOST is operated by the University of Sydney with 
support from the Australian Research Council and the Science Foundation for Physics within the University of Sydney.} with a 
resolution of 43\arcsec\,$\times$43\arcsec\,cosec\,\textbar Declination\textbar\,. For our analysis purpose, we retrieved the 
original processed image for this region's Galactic coordinates (resolution\,$\sim$\,55.84\arcsec$\times$43\arcsec\,) from the 
website. The radio images are used to obtain the physical parameters and examine the morphology of the ionized gas in the 
region (see Section \ref{section_RadioMorphology}).

\subsection{Other Archival Data Sets}
\label{section_ArchivalData}

\subsubsection{\textit{Spitzer} mid-infrared observations}
\label{section_SpitzerData}

Archival MIR observations of this region, obtained using the \textit{Spitzer} Space Telescope under the 
Galactic Legacy Infrared Midplane Survey Extraordinaire (GLIMPSE) program \citep{benjamin03, churchwell09}, were retrieved 
with the help of the InfraRed Science Archive\footnote{This research has made use of the NASA/ IPAC Infrared Science Archive, 
which is operated by the Jet Propulsion Laboratory, California Institute of Technology, under contract with the National 
Aeronautics and Space Administration.} (the Spring '07 Archive more complete catalogue). GLIMPSE observations were taken using 
the InfraRed Array Camera (IRAC) in 3.6, 
4.5, 5.8, and 8.0 $\mu$m bands. The final image cutouts (pixel scale of 0.6\arcsec\,) as well as the final catalogues were 
downloaded. While the images are further used to analyse features in the region, the photometric catalogue, in conjunction 
with the NIR catalogue (Section \ref{section_NIR_Photometry}) was used to identify the YSOs in the region.

\subsubsection{\textit{Herschel} far-infrared observations} 
\label{section_HerschelData}

This region has been observed at FIR wavelengths, in the range 70-500\,\micron\,, using the instruments 
Photodetector Array Camera and Spectrometer \citep[PACS;][]{poglitsch10} and Spectral and Photometric Imaging
Receiver 
\citep[SPIRE;][]{griffin10} on the 3.5\,m 
\textit{Herschel} Space Observatory \citep{pilbratt10}, as a part of the Proposal ID 
\textquotedblleft KPOT\_smolinar\_1\textquotedblright\, \citep{molinari10}. For our analyses, we obtained the PACS 70\,\micron\, 
and 160\,\micron\, 
level2\_5 MADmap images \citep{cantalupo10}; and the SPIRE 250\,\micron\,, 350\,\micron\,, and 500\,\micron\, level2\_5 extended 
source (\textquotedblleft extdPxW\textquotedblright\,) map products using the \textit{Herschel} Science 
Archive\footnote{http://www.cosmos.esa.int/web/herschel/science-archive}. 
The 70\,\micron, 160\,\micron\,, 250\,\micron\,, 350\,\micron\,, and 500\,\micron\, images have pixel scales of 3.2\arcsec\,, 
6.4\arcsec\,, 6\arcsec\,, 
10\arcsec\,, and 14\arcsec\,, respectively, with resolutions varying from $\sim$\,5.5\arcsec\,--36\arcsec\,. While the PACS 
image obtained had the surface brightness unit of Jy\,pixel$^{-1}$, the SPIRE images were in the units of MJy\,sr$^{-1}$. 
Detailed information about the data products is 
provided on the \textit{Herschel} site\footnote{http://www.cosmos.esa.int/web/herschel/data-products-overview}. We use the 
FIR data to examine the physical conditions of the region.

\section{Stellar Population in the region}
\label{section_StellarPopulation}

\subsection{Identification of YSOs} 
\label{section_YSO} 

\begin{figure*}
\centering
\subfigure
{
\includegraphics[scale=0.5]{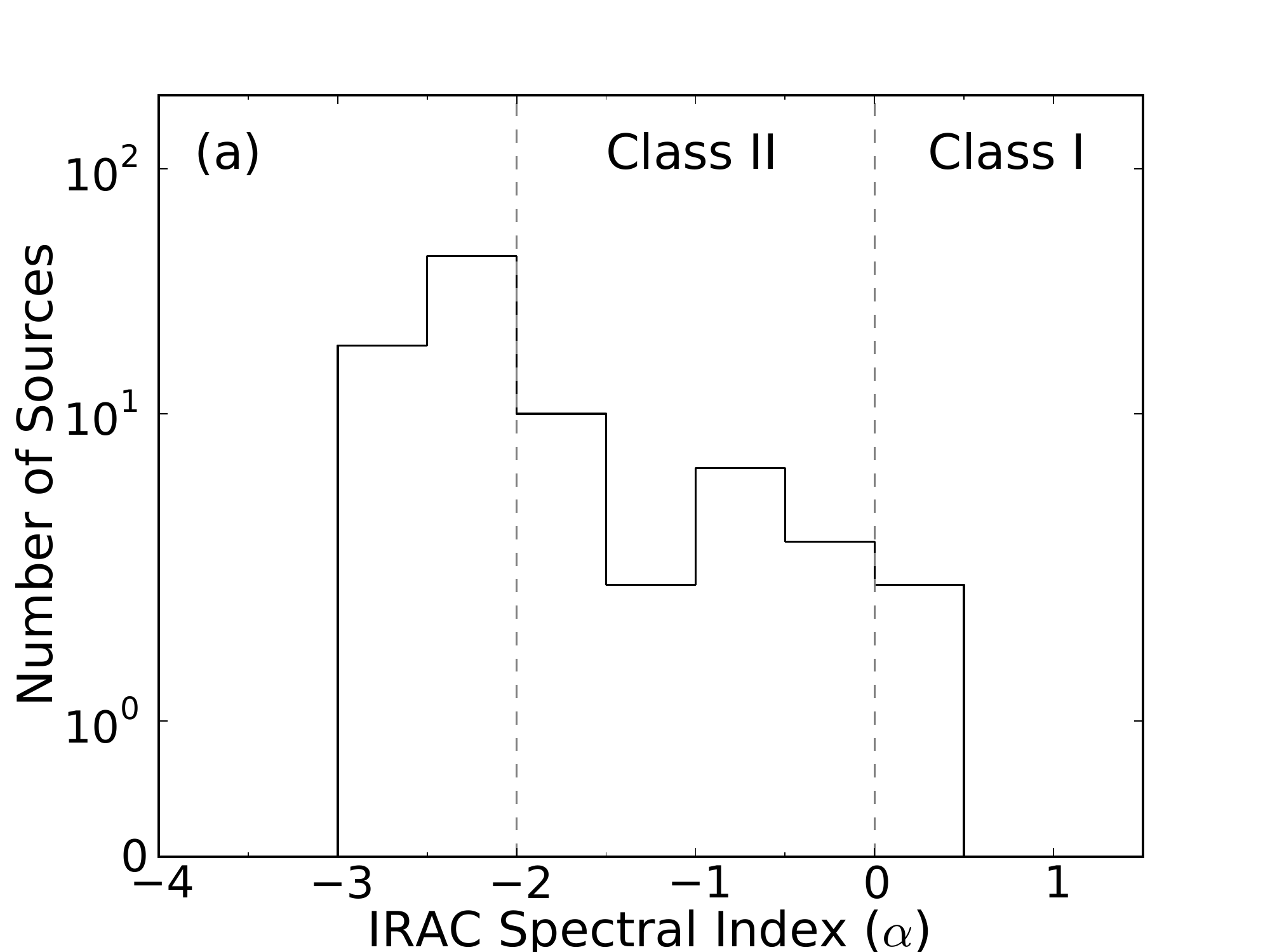}
\label{fig_YSO_SpectralIndex}
}
\subfigure
{
\includegraphics[scale=0.5]{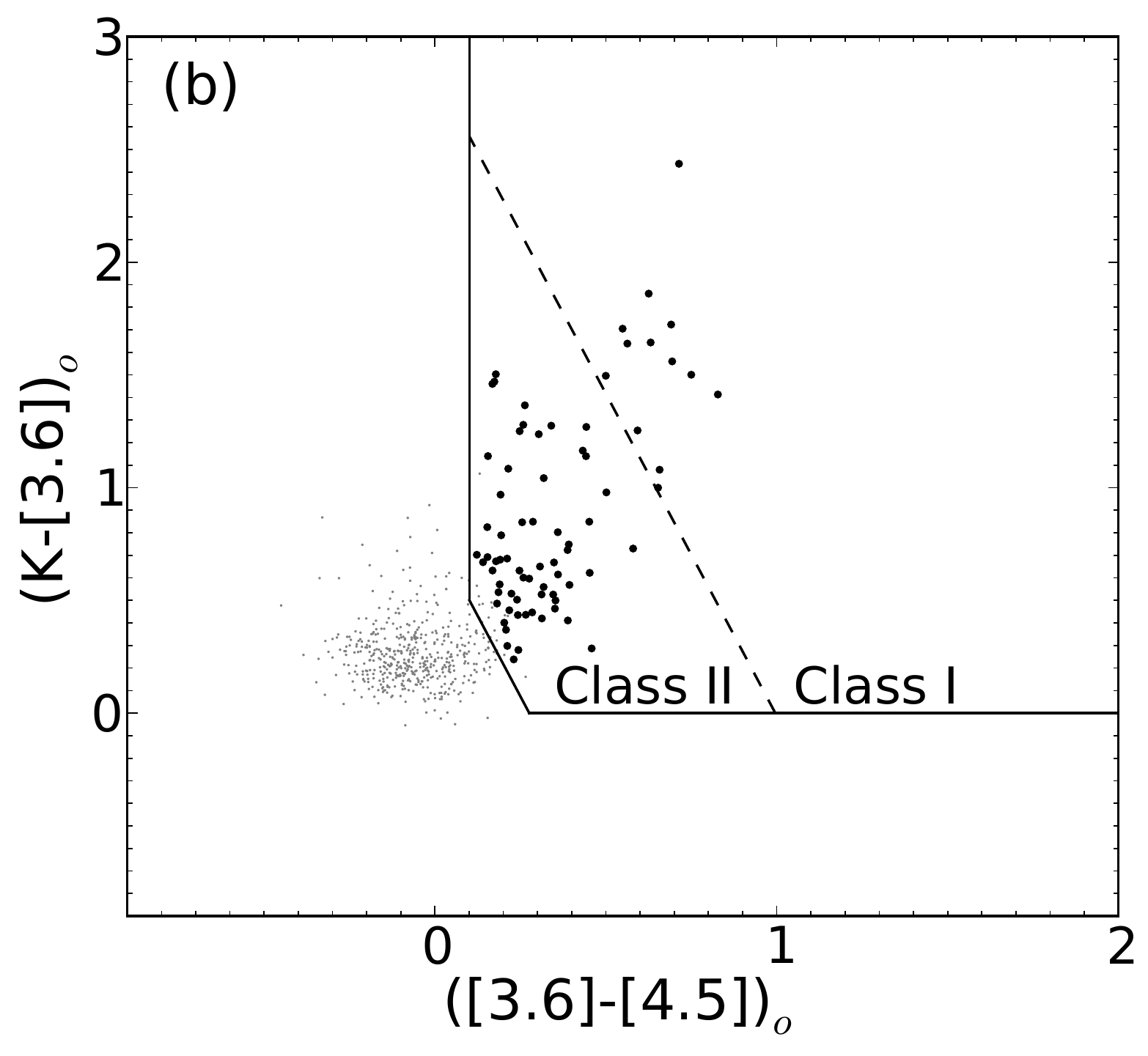}
\label{fig_YSO_Gutermuth} 
}
\subfigure
{
\includegraphics[scale=0.5]{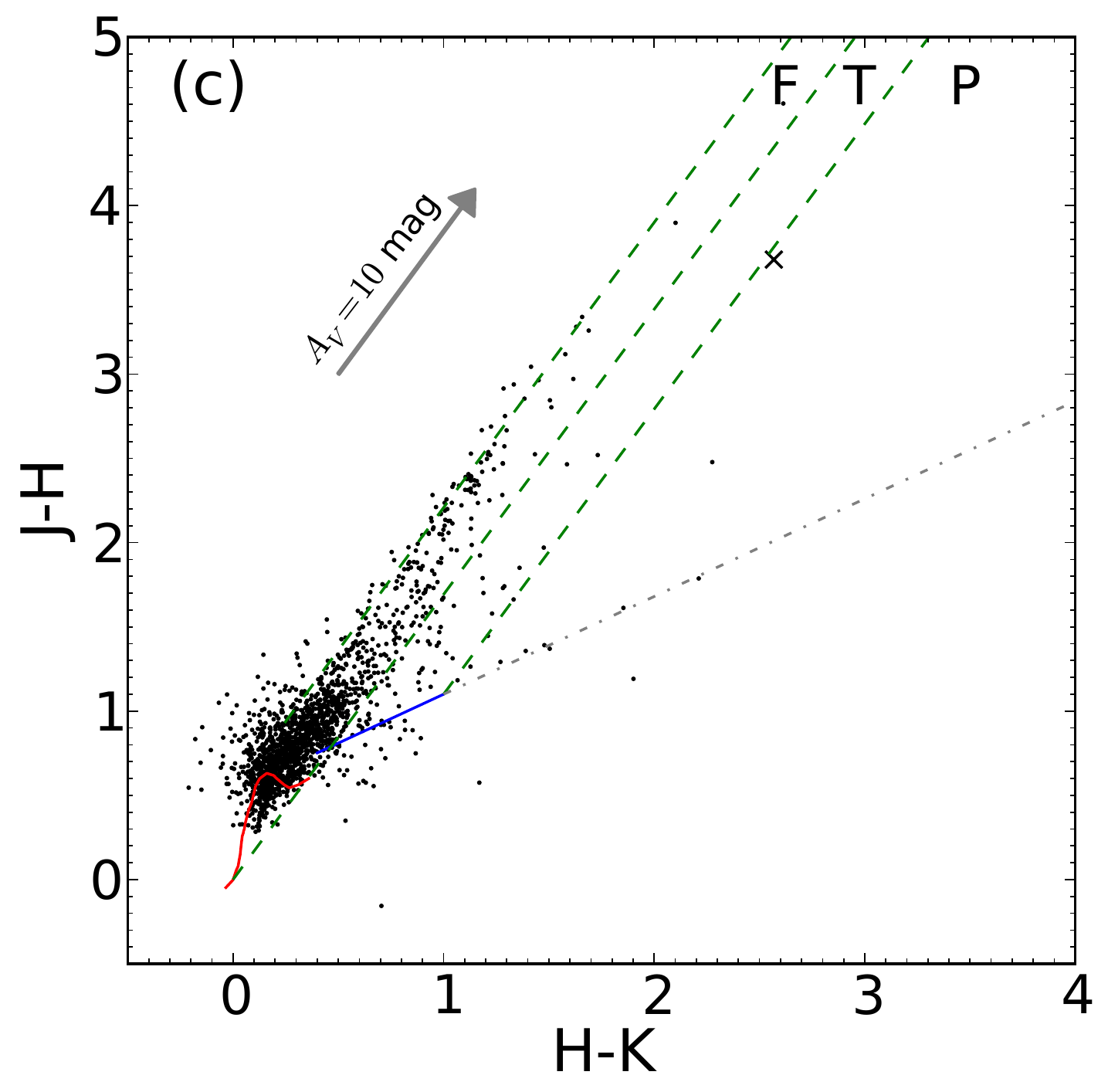}
\label{fig_YSO_NIR_CCD} 
}
\caption{(a) Histogram of the IRAC spectral indices. The limits for Class\,I and Class\,II sources have been marked by dashed 
grey vertical lines. (b) NIR-MIR CCD using the procedure of \citet{gutermuth09}. The regions of Class\,I and Class\,II sources 
have been labelled. The identified YSOs have been shown with black solid circles. (c) NIR CCD in CIT system. The red curve and 
the blue line show the dwarf locus \citep{bessell88} and the CTTS locus \citep{meyer97}, respectively. The grey dot-dashed line 
is the extension of CTTS locus. Three parallel slanted dashed lines mark the reddening vectors, drawn using extinction laws from 
\citet{cohen81}. 
Three separate regions, \textquoteleft F\textquoteright\,, \textquoteleft T\textquoteright\,, \textquoteleft P\textquoteright\, 
have been labelled on the plot. The source marked with a cross in \textquoteleft P\textquoteright\, region is a high-mass YSO 
(see text).}  
\label{fig_YSO_Identification} 
\end{figure*}

The NIR and MIR photometric catalogues from IRSF (Section \ref{section_NIR_Photometry}) and GLIMPSE 
(Section \ref{section_SpitzerData}), respectively, were cross-matched within 0.6\arcsec\, matching radius and collated 
to obtain a combined photometric catalogue. Thereafter, the following set of steps were followed for the identification 
of the YSOs \citep[similar to][]{mallick13} : 
\begin{enumerate}
\item 
First, the YSOs were identified using their MIR magnitudes. The sources with detections in all four IRAC bands with 
errors $\leq$\,0.15\,mag were used here. Using simple linear regression, the IRAC spectral index 
\citep[$\alpha_{IRAC} = d\log (\lambda F_{\lambda})/d\log (\lambda)$;][]{lada87} was calculated for each source (in the 
wavelength range 3.6-8.0\,\micron), followed 
by their classification into Class\,I and Class\,II categories using the limits from \citet{chavarria08} 
($\alpha_{IRAC} > 0$ for Class\,I, and $-2 \leq \alpha_{IRAC} \leq 0$ for Class\,II; see Fig. \ref{fig_YSO_SpectralIndex}). 
\item 
All sources need not have detections at 5.8 and 8.0\,$\mu$m, but might have good quality detections in NIR bands. To 
identify the YSOs from amongst such sources, we use a combination of $H, K_s$, 3.6\,$\mu$m, and 4.5\,$\mu$m bands 
following the procedure of \citet{gutermuth09}. Again, only those sources whose 3.6 and 4.5\,$\mu$m magnitude errors 
are $\leq$ 0.15 are used here. The YSOs were identified from their location in the dereddened 
- using the colour excess ratios from \citet[][]{flaherty07} - 
\textquotedblleft $K_s$-[3.6]\textquotedblright\, versus \textquotedblleft [3.6]-[4.5]\textquotedblright\, 
colour-colour diagram (CCD), as is shown in Fig. \ref{fig_YSO_Gutermuth}. 
\item 
Additional YSOs were identified using the $J-H/H-K$ CCD (Fig. \ref{fig_YSO_NIR_CCD}) according to the following procedure. In 
Fig. \ref{fig_YSO_NIR_CCD}, the red solid curve marks the dwarf locus from \citet{bessell88} and the blue solid line marks the 
Classical T Tauri Stars (CTTS) locus from \citet{meyer97}. All the loci curves as well as sources' colours were converted 
to CIT (California Institute of Technology) photometric system \citep[using][]{carpenter01} for this analysis. 
The slanted dashed lines are the reddening vectors, drawn using the reddening laws of \citet{cohen81} for the CIT photometric 
system. In this CCD, three separate 
regions have been marked, similar to \citet{ojha04a,ojha04b}. The sources in \textquoteleft T\textquoteright\, and 
\textquoteleft P\textquoteright\, regions are taken to be Class\,II sources \citep{lada92}, as they exhibit IR-excess 
emission. In the \textquoteleft P\textquoteright\, region, since there could be a slight overlap between Herbig\,Ae/Be stars 
and Class\,II sources \citep{hillenbrand92}, we conservatively took only those sources which were above the CTTS 
locus extended into this region \citep[similar to][]{mallick14}. 
It should be noted that sources in the \textquoteleft T\textquoteright\, region could also contain a few Class\,III 
sources with small IR-excess. 
\end{enumerate}
Finally, we merged the sources identified in each step. An overlapping source might have different identifications in 
different steps. Thus, in the final list, the class of a YSO was taken as that in which it was identified as first in 
the above order of steps. A total of 7 Class\,I and 133 Class\,II sources were obtained (for a FoV of 
$\sim$\,5.5\arcmin$\times$5.5\arcmin\, encompassing the molecular cloud, as marked in 
Fig. \ref{fig_Introduction_ColourComposite}) in the final YSO catalogue, which is given in Table \ref{table_YSOs}.

\subsection{Spectral Energy Distribution of YSOs}
\label{section_SED}

SED modelling was carried out for (a subset of) YSOs to get estimates of their physical parameters. The grid of YSO models 
from \citet{robitaille06} - implemented in the online SED fitting tool of \citet{robitaille07} - was used for this purpose. 
The basic model consists of a pre-main sequence (PMS) star surrounded by a flared accretion disk having a rotationally 
flattened envelope with cavities carved out by a bipolar outflow. A total of 200000 SED models are computed in a 14 dimensional
parameter space (covering properties of the central source, the infalling envelope, and the disk), using the radiation transfer 
code of \citet{whitney03a,whitney03b}. The online fitting tool attempts to fit the available SED models to the data, 
characterising each fitting by a $\chi^{2}$ parameter. The distance range and the interstellar visual extinction (A$_{V}$) are 
free parameters whose range has to be specified by the user. Accounting for the uncertainties, we adopted a large distance range
of 3.4 to 3.8\,kpc for our sources. 
From extinction calculations (see Section \ref{section_ClusterAndExtinction}) we find that almost all non-YSO sources had 
A$_V \leq$\,20\,mag, and thus we specified the range of interstellar visual extinction A$_{V}$ as 1-20 mag. 

Since the number of SED models is very large, spanning a wide range of parameter space, the models fitting each source can 
only be constrained by increasing the number of data points used, and having data points to sufficiently cover the entire 
wavelength range of fitting. For this reason, we choose the sources with photometry in at least all four IRAC bands 
for SED modelling. 
However, in addition to those selected using this criterion, we also carried out SED analysis for the prominent sources 
associated with the central nebula (even though two of them lacked 8.0\,\micron\, photometry). 
We note that photometry from \emph{Herschel} images is not useful here as none of the YSOs have counterparts at 
those wavelengths, which in part is due to poor resolution.

For each source, the SED fitting tool - besides giving the best fit model - also gives a set of well fit models ranked by 
their $\chi^{2}$ values as a measure of their relative \textquotedblleft goodness-of-fit\textquotedblright. Following a method 
similar to \citet{robitaille07}, we consider only those models for further calculation which satisfied the following criterion :
\begin{equation} 
\label{equation_chi} 
\chi^{2} - \chi^{2}_{min} < 3 \ \rm{(per\ data\ point)} . 
\end{equation} 
As elucidated in \citet{robitaille07}, though this criterion is based on visual examination of SED plots and has no rigid 
mathematical backing, a stricter criterion might lead to over-interpretation. For each parameter, the weighted mean value 
and standard deviation were calculated using the models which satisfied Equation (\ref{equation_chi}). The inverse of the 
respective $\chi^{2}$ was taken as the weight for each model \citep[similar to][]{grave09}. Table \ref{table_SED} gives our 
SED modelling results, listing the physical parameters :  
age of the central source (\textit{t$_{*}$}), mass of the central source, disk mass ($M_{disk}$), disk accretion rate 
($\dot M_{disk}$), envelope mass ($M_{env}$), temperature of the central source ($T_*$), total system luminosity ($L_{total}$), 
interstellar visual extinction (\textit{A$_{V}$}), and the $\chi^{2}_{min}$ per data point. Errors for some parameters tend to 
be large as we are dealing with a large parameter space while we have very few data points to constrain the number of models. 
If a source was simply fit better as a star with high interstellar extinction, it has not been included in the table. 
Thus finally, SED results are given for a total of 3 Class\,I sources, 24 Class\,II sources (including 2 central sources), and 
one extra central source. 

\begin{figure*}
\centering
\subfigure
{
\includegraphics[scale=0.7]{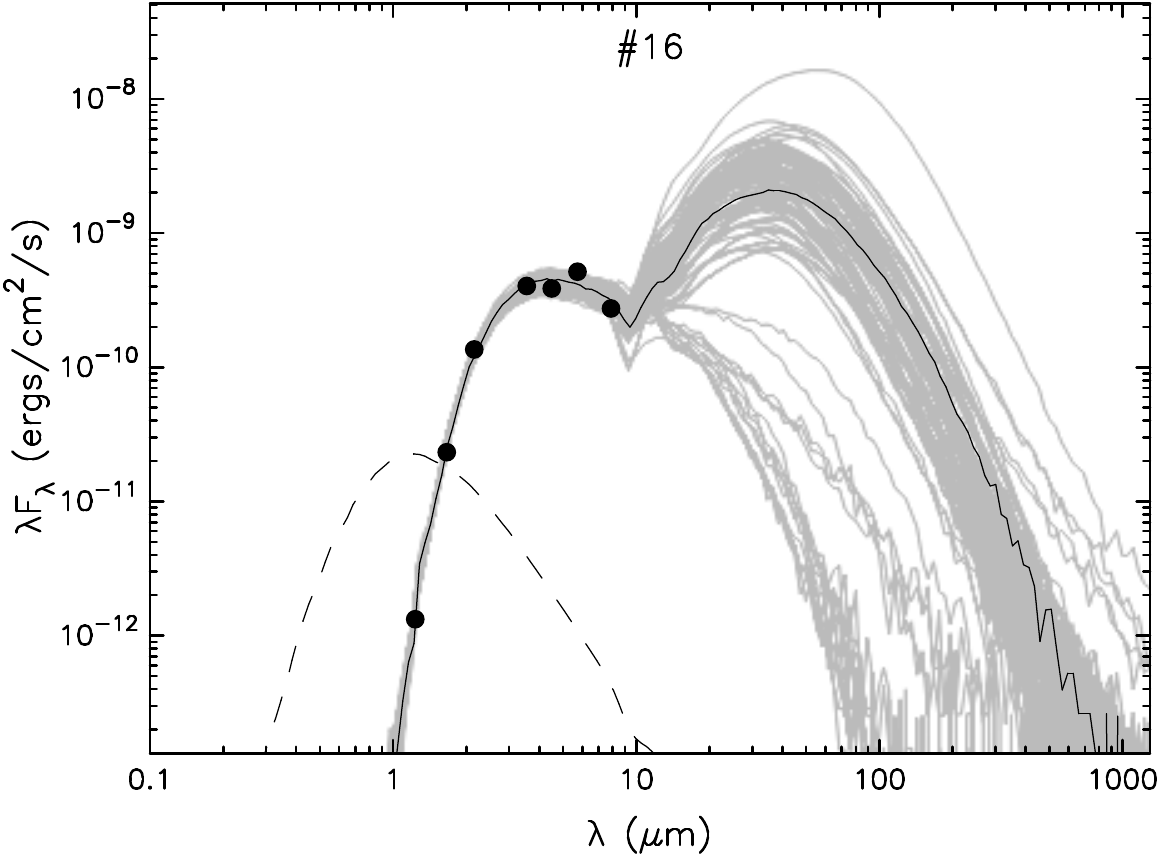}
}
\caption{SED fitting using the online tool of \citet{robitaille07} for source \#16 (a Class\,II source) in Table \ref{table_SED}.
The black dots mark the data points. The solid black curve is the best fitted model, while the grey curves denote the 
subsequent good fits for $\chi^2 - \chi^2_{min} (per~data~point) < 3$. The dashed curve is the photosphere (in the presence of 
interstellar extinction, but absence of circumstellar dust) of the central source for the best-fit model.}  
\label{fig_SED_Source17} 
\end{figure*}

Even though the statistics for the YSOs (i.e. Class\,I and Class\,II sources) is not very significant, we can still use them to 
get an idea of the physical parameters of the stellar sources forming in this region. As can be seen from Table \ref{table_SED}, 
there appears to be a considerable age dispersion, with the ages ranging from $\sim$ 0.05\,Myr to 0.5\,Myr for most of the 
sources. 5 YSOs even have ages $>$\,1\,Myr. This is suggestive of ongoing star formation. 
All the YSOs analysed here yield masses $>$\,2\,\msun, and five of them $>$\,6\,\msun. One of the YSOs (\#16) appears 
to be a high-mass star of $\sim$\,9.6\,\msun, and is embedded in the nebular emission associated with this region 
(see Section \ref{section_Morphology}). The SED plot for this source is shown in Fig. \ref{fig_SED_Source17}. 
\citet{grave09} had also done SED analysis for what they mention as an embedded point source in this IR nebula with 
$JHK$ and four IRAC bands. Since among the possible embedded sources in the central part of this IR nebula, only this source 
(\#16) had all 7 magnitudes, most likely their \textquoteleft 16148-5011mms2near\textquoteright\, refers to this source itself. 
\citet{grave09} calculated the age and mass of this source as 4.2$\pm$0.3 ($\log t_*$) and 11$\pm$1\,\msun, respectively. 
Though the mass estimate appears consistent (within error limits) with our results, their age estimate is much lower. It is 
probable that this could be because they use different distance estimates and 1.2\,mm fluxes from literature.
This source is also classified as a YSO by 
\citet[called \textquotedblleft G333.0494$+$00.0324B\textquotedblright\,]{lumsden13}.
Another source (\#28) associated with the nebula appears to be a high-mass stellar object, with age\,$>$\,1\,Myr, though it is 
not classified as a YSO in our analysis. 

It should be noted that the SED results are only representative of the actual values, as an empirically consistent fit might 
not be the correct fit. A deliberate sampling bias of the huge parameter space, to reduce computational time, could give 
rise to pseudotrends in the results. The results are also contingent upon the validity of assumed evolutionary tracks from 
literature. Most importantly, the models are for individual sources, and could be misleading in cases of stellar multiplicity. 
The caveats are dealt in detail in \citet{robitaille08}.

\subsection{Luminosity Function} 
\label{section_LF} 

\begin{figure*}
\centering
\subfigure
{
\includegraphics[scale=0.45]{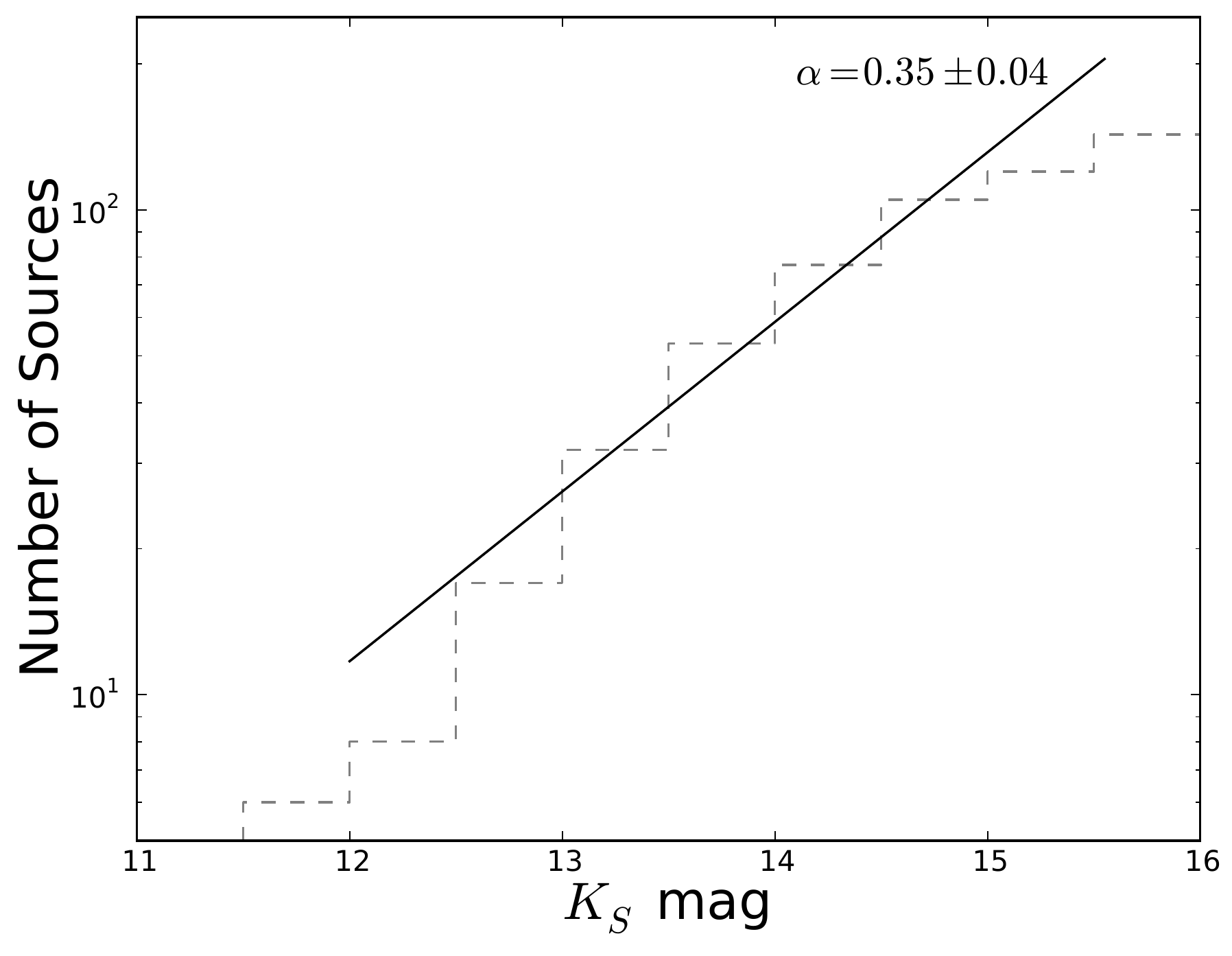}
}
\caption{The grey dashed histogram shows the cumulative KLF for the YSOs. The black straight line is the fit in 
[12,15.5]\,mag range, whose slope is given by $\alpha$.}  
\label{fig_KLF_YSOs} 
\end{figure*}

The slope of the $K_s$-band luminosity function (KLF) can serve as an indicator of the age of a stellar cluster 
\citep{zinnecker93,lada95,vig14}. 
If we were to assume that the mass function and the mass-luminosity relation for a (coeval) stellar cluster are power laws, 
i.e. they are of the form $dN(\log m_*) \propto m_*^{-\gamma} d\log m_*$ and $L_K \propto m_*^B$, then it can be shown that 
the slope of the KLF will be of the form : 
\begin{equation}
\alpha = \gamma/2.5 B
\end{equation}
\citep{lada93,megeath96}. 
First of all, we try to estimate the KLF slope. We only 
consider the YSOs here, as opposed to all the observed sources, as they will be much less affected by any field star contamination. 
Our \emph{K}-band 100\% completeness limit is upto 14\,mag, and thus completeness correction was implemented in the  
(0.5\,mag sized) bins after this limit. Thereafter, the (cumulative) KLF of the YSOs was constructed, and is shown in 
Fig. \ref{fig_KLF_YSOs}. 
The fit to the histogram is also shown, and its slope ($d\log N/dm_K$) in [12,15.5] mag range is calculated to be ($\alpha =$)
0.35$\pm$0.04 \citep[the slope will be same for cumulative and differential KLFs here, see][]{lada93}. 
As for the slope of mass-luminosity relation, if we adopt the value of $B=2$ (which can be shown to be the approximate value 
for O-F main sequence stars; \citealt{lada93}), then the mass function slope comes out to be ($\gamma =$)1.75$\pm$0.20 (a value 
slightly steeper than the Salpeter slope of 1.35).  

\begin{figure*}
\centering
\subfigure
{
\includegraphics[scale=0.5]{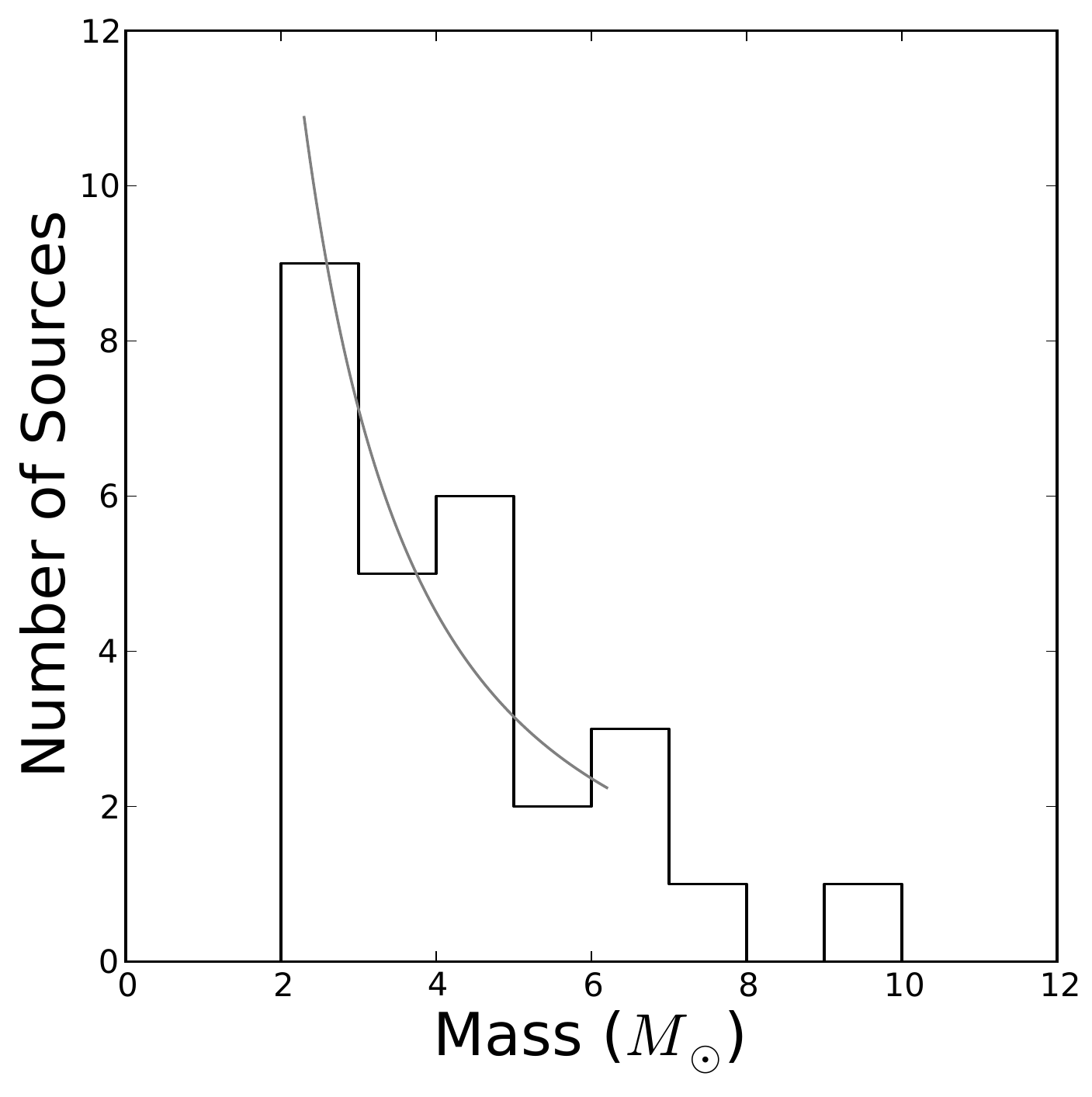}
}
\caption{Histogram of stellar masses obtained from the SED fitting (Table \ref{table_SED}). The grey curve shows the power 
law fit in the intermediate mass range.}  
\label{fig_MassFunction} 
\end{figure*}

Alternatively, we could get an estimate of the mass function slope using the SED fitting results. Since most of our sources 
are of intermediate mass in 2-6\,\msun\, range (see Table \ref{table_SED}), we consider only this range to estimate the 
slope $\gamma$. Fig. \ref{fig_MassFunction} shows the mass histogram (for the YSOs). Assuming that the 
star formation is strictly coeval, the mass function slope is given by $\gamma = -(d\log N/d\log m_*)$ \citep{massey98}, and thus 
we calculate $\gamma \sim 1.59 \pm 0.70$ for our case. The grey curve in Fig. \ref{fig_MassFunction} shows the fitted function for 
2-6\,\msun. This is consistent with $\gamma$ obtained above. The major source of uncertainty here is the sparse statistics, and 
thus possible incompleteness in the mass bins. 

A similar set of values (both $\alpha$ and $\gamma$) was found by \citet{balog04} for the 1\,Myr old NGC\,7538 cluster 
\citep[distance\,$\sim$\,2.65\,kpc;][]{mallick14}. $\alpha$ also appears consistent with that derived for the Orion molecular 
clouds \citep[$\sim$\,0.37-0.38, age\,$\sim$\,1\,Myr;][]{lada95,lada91}. 
However, other embedded clusters in literature, such as  
the NGC\,1893 cluster \citep[$\sim$\,0.34$\pm$0.07, distance\,$\sim$\,3.25\,kpc;][]{sharma07}, 
the Tr\,14-16 clusters in Carina nebula \citep[$\sim$\,0.30-0.37, distance\,$\sim$\,2.5\,kpc;][]{sanchawala07}, 
and the IRAS\,06055+2039 cluster \citep[$\sim$\,0.43$\pm$0.09, distance\,$\sim$\,2.6\,kpc;][]{tej06} do exhibit slightly 
larger cluster ages, upto $\sim$\,4\,Myr. Hence it appears that 1\,Myr should be the lower age limit of the cluster.

\subsection{Mass Spectrum}
\label{section_MassSpectrum} 

\begin{figure*}
\centering
\subfigure
{
\includegraphics[scale=0.5]{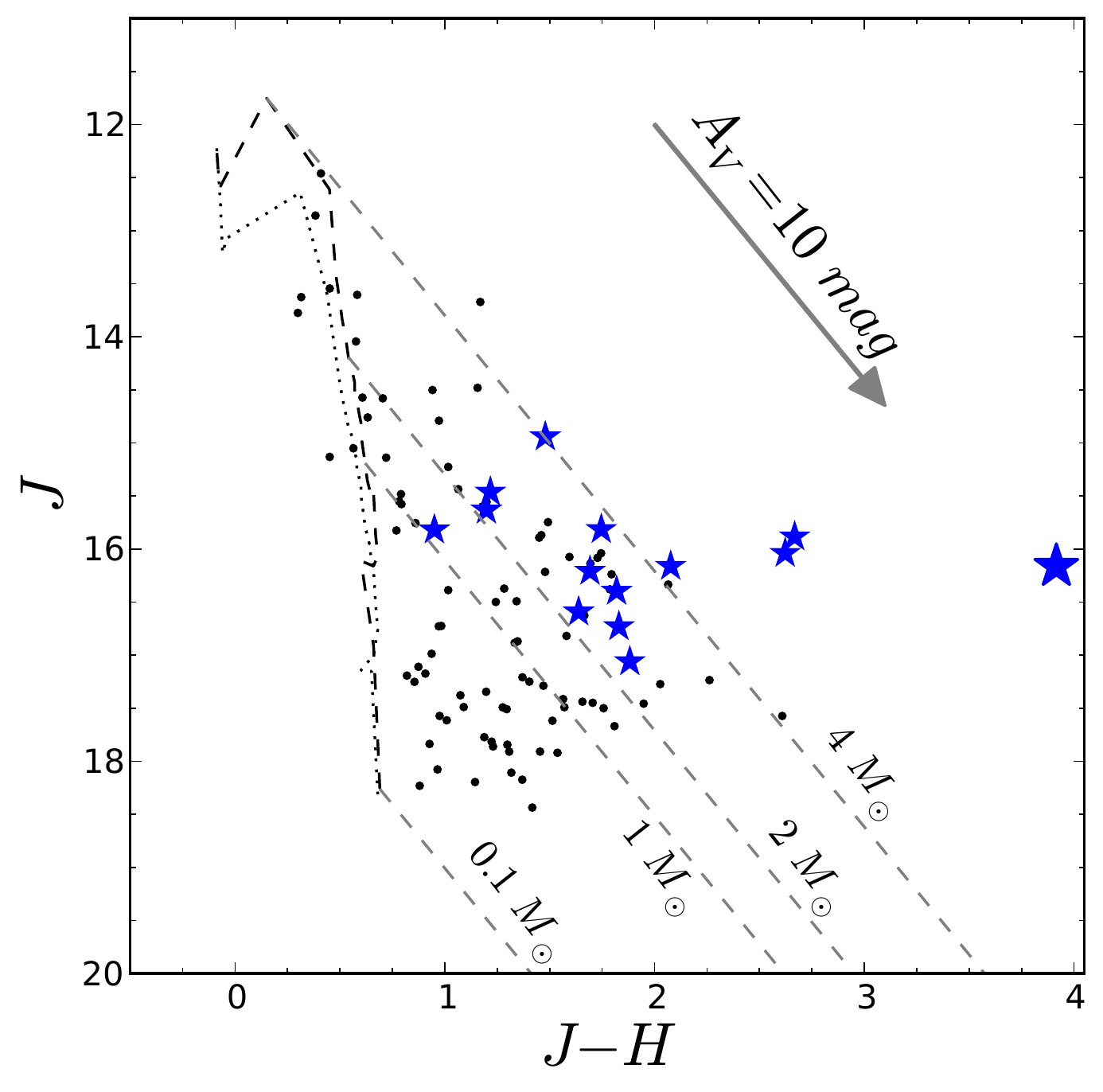}
}
\caption{$J$/$J-H$ colour-magnitude diagram for the YSOs with at least $J$ and $H$ band detections. 
The dashed and dotted black curves plot the 1\,Myr and 2\,Myr PMS isochrones, respectively, 
from \citet{siess00}. The reddening vectors (parallel grey dashed lines) for the 1\,Myr isochrone 
are drawn at 0.1, 1, 2, and 4\,\msun. The blue star symbols are the sources for which SED analysis was done.}  
\label{fig_MassSpectrum} 
\end{figure*}

In addition to the SED, we also use the $J/J-H$ colour-magnitude diagram (CMD) to get an estimate of the mass range of the YSOs. 
We avoid using a CMD involving the $K$-band magnitude as this band is the most affected by the NIR excess flux arising 
from circumstellar material, which in turn can lead to brightening, and thus erroneous mass estimate, of the sources.  
Fig. \ref{fig_MassSpectrum} shows the $J/J-H$ CMD for the YSOs with at least $J$ and $H$ band detections. 
The 1\,Myr PMS isochrone, along with the 2\,Myr isochrone for reference, from \citet{siess00} has been shown on the image. 
For the 1\,Myr PMS isochrone, reddening vectors are shown for 0.1, 1, 2, and 4\,\msun\,. 
As can be seen, all but one of the YSOs for which the SED analysis has been done (marked with blue stars) lie in the mass range 
$\ga$\,2\,\msun\,, matching well with our SED results. In general, the sources lie in the mass range $\sim$\,0.1-4\,\msun, or, 
put differently, the observations probe stellar objects upto the 0.1\,\msun\, limit. 
The (blue star) source at the far-right end of the diagram is the 9.6\,\msun\, high-mass source from 
SED analysis. The wide variation in the colours of YSOs is probably an indication of variable extinction as well as 
different evolutionary stages of the sources. Major causes of uncertainty here are : uncertainty in distance estimate which is 
used to obtain the apparent magnitude (for the isochrones), and unresolved (especially since the source distance of 
$\sim$\, 3.6\,kpc is much larger than most of the previously studied clusters) binarity.  
Additionally, it should be kept in mind that, in general, a particular PMS model will introduce its own systematic error.

\section{Morphology of the Region}
\label{section_Morphology}

\subsection{Cluster Analysis and Extinction mapping}
\label{section_ClusterAndExtinction} 

\begin{figure*}
\centering
\subfigure
{
\includegraphics[scale=0.5]{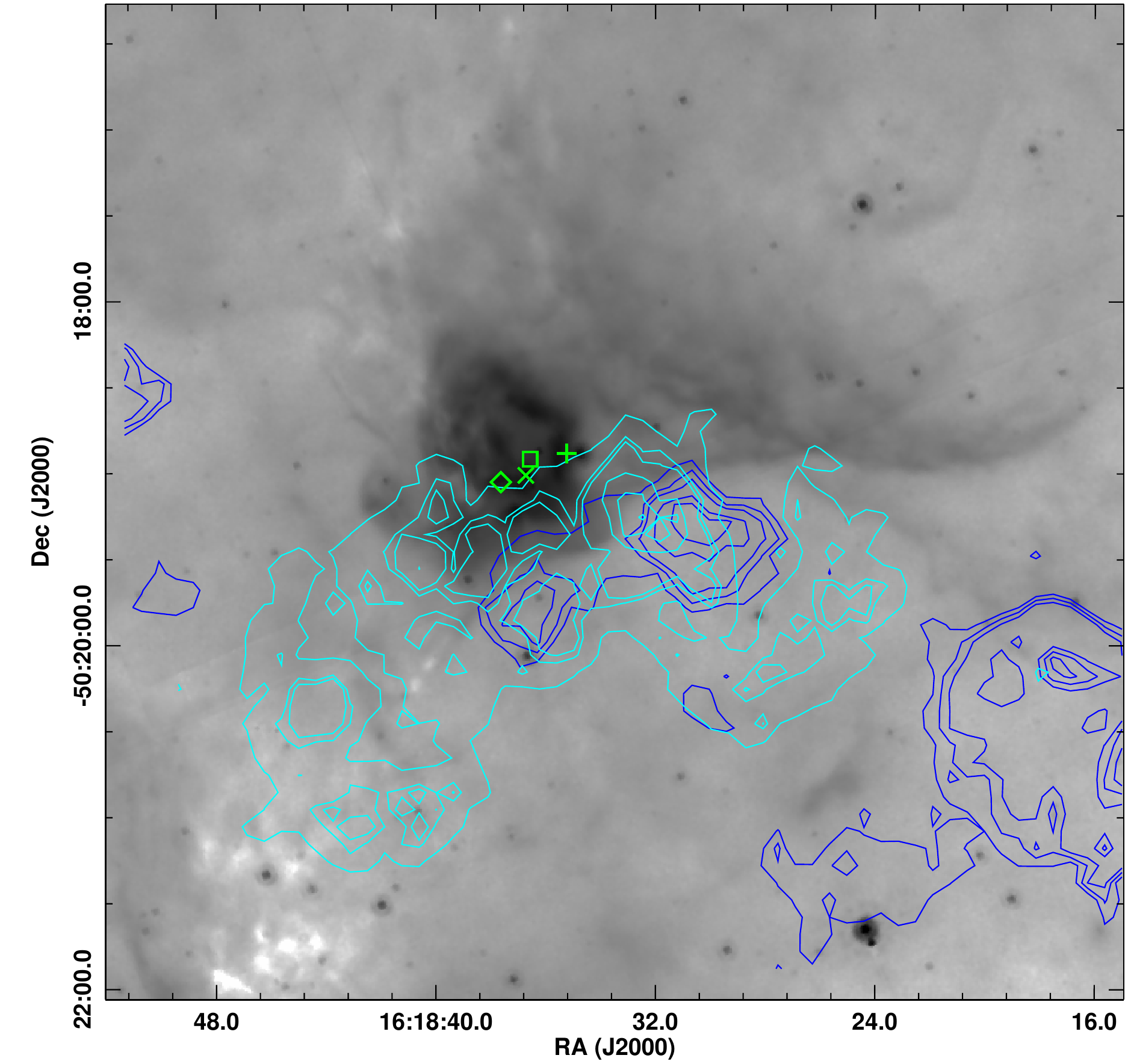}
}
\caption{\textit{Spitzer} 8.0\,$\mu$m image of the region with overlaid surface density contours (at 5, 5.75, 6, 7, 
7.5, and 8 YSOs pc$^{-2}$) in cyan, and visual extinction contours (at A$_V$\,=\,4, 4.5, 5, 6, and 6.5\,mag) in blue.
Green plus symbol marks the IRAS catalogue position of IRAS\,16148-5011, green cross the high-mass source (\#16 
from Table \ref{table_SED}), while the diamond and box symbols mark the millimeter and MSX peaks, respectively, 
from \citet{molinari08}.}  
\label{fig_NNdensity_extinction} 
\end{figure*}

Fig. \ref{fig_NNdensity_extinction} shows the \textit{Spitzer} 8.0\,$\mu$m image with overlaid surface density contours (in 
cyan) as well as visual extinction contours (in blue). 
The high-mass source in this region (see Section \ref{section_SED}), as well as the millimeter (mm) and MSX peaks from 
\citet[16148-5011\,MM\,1 and 16148-5011\,A from their Tables\,2 and 3, respectively]{molinari08} have also been marked on the 
image. The marked mm peak is almost coincident with the (mm) peak from \citet[16148-5011\,Clump\,2 from their Table\,2]{beltran06}. 
The surface density and extinction contours were calculated as follows. 
The surface density analysis was carried out using the nearest-neighbour (NN) method \citep{casertano85} to discern the YSO 
clusterings in the region. We chose 20\,NN, similar to \citet{schmeja08, schmeja11, mallick13}. 
The extinction in the region was estimated by constructing the extinction map with the help of the NIR photometric data. 
We use the NIR CCD (see Fig. \ref{fig_YSO_NIR_CCD}) for this purpose. In this NIR CCD, the \textquoteleft F\textquoteright\, 
region mostly contains the main-sequence field sources along with a few probable Class\,III sources. Since these are sources which 
have 
almost lost their circumstellar material, any extinction they exhibit will come from interstellar - rather than circumstellar - 
material. The sources from this \textquoteleft F\textquoteright\, region, which were not identified as a YSO in 
Section \ref{section_YSO}, were therefore selected. 
Subsequently, we used a method similar to the Near-IR-Colour-Excess method of \citet{lada94}, but here the $(H-K)$ colour excess 
was estimated by dereddening the sources - along the reddening vector - to the low-mass end (turnover onwards) of the dwarf 
locus (which was approximated as a straight line). $A_V$ was then calculated using the reddening laws 
of \citet{cohen81}. After we obtained the visual extinction $A_V$ for each source, an extinction map of the region was made by 
using the NN method, where the extinction at each grid point on the map is the median (because median rejects outliers) of $A_V$ 
of 20\,NN. 
It is possible that the extinction could be slightly underestimated because, due to the large distance to this region, 
a significant fraction of the detected sources could be foreground sources, specially towards the center of the nebula.

In Fig. \ref{fig_NNdensity_extinction}, 
the surface density contours are drawn at 5, 5.75, 6, 7, 7.5, and 8 YSOs\,pc$^{-2}$, while the extinction contours have been drawn 
at A$_V =$ 4, 4.5, 5, 6, and 6.5\,mag levels. As can be seen on the image, both the surface density as well as extinction contours 
are in the 
southern portion of the nebular emission and appear to be along the sharp boundary of the nebula. The highest surface density 
contour levels coincide with the highest extinction levels. Though we would have expected to see high interstellar extinction  
along the main body of the nebula, this is not so possibly because the NIR observations are not deep enough to detect stars from 
behind the nebula and thus the extinction of the nebula should be higher than the highest extinction contour level here 
(it is later estimated to be A$_V >$\,20\,mag; see Section \ref{section_HerschelResults}) 
It should also be noted that the cluster detected here is in the southern part of the nebula also coincident with extended radio 
continuum emission (see Section \ref{section_RadioMorphology}).

\subsection{Radio Morphology} 
\label{section_RadioMorphology}

\begin{figure*}
\centering
\subfigure
{
\includegraphics[scale=0.5]{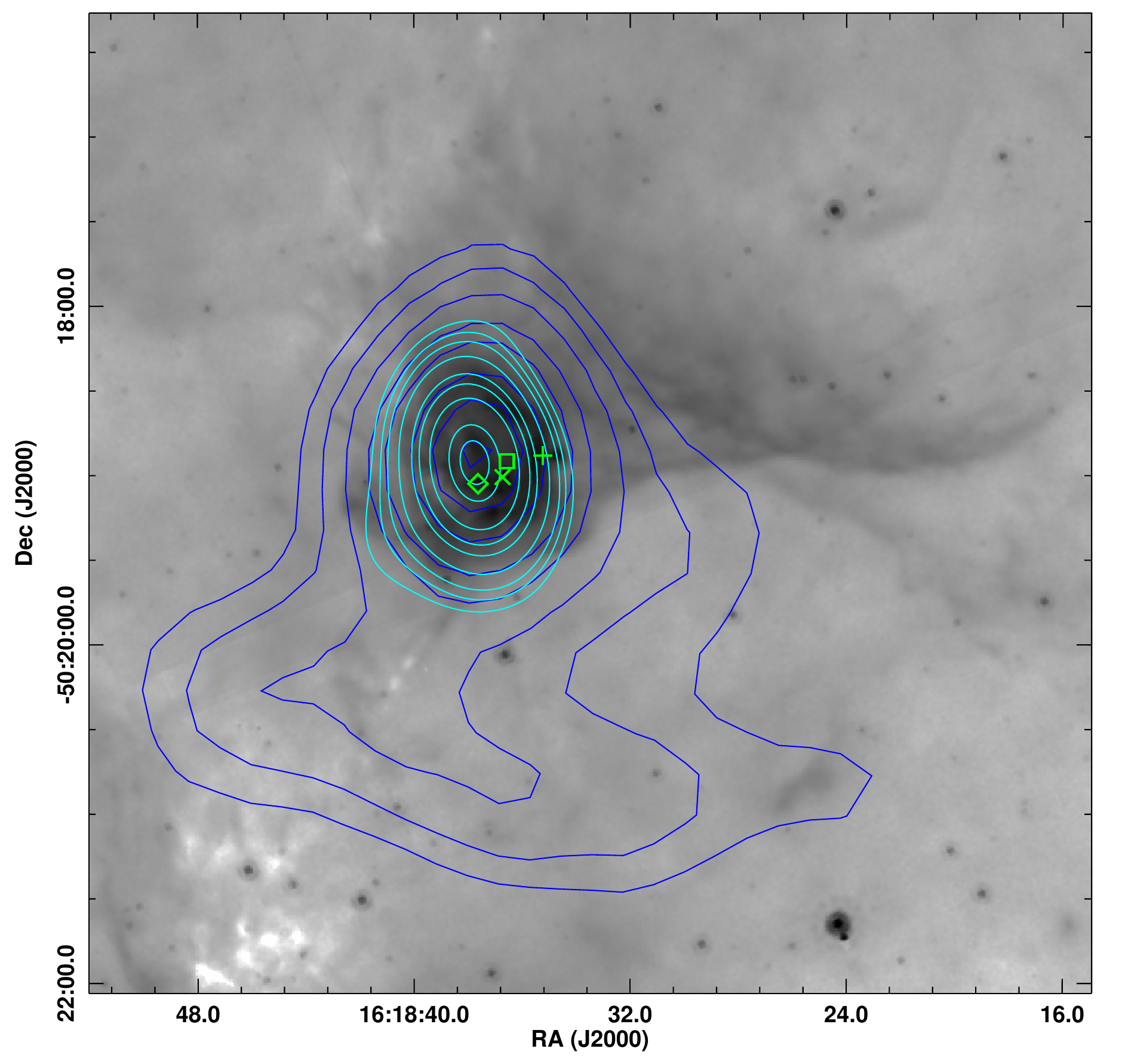}
}
\caption{\textit{Spitzer} 8.0\,$\mu$m image of the region with overlaid 1280\,MHz contours 
(at 3, 4, 5, 7, 10, 12, 15, 20, and 22\,$\sigma$, where $\sigma \sim$\,8.85\,mJy; resolution\,$\sim$\,56\arcsec$\times$35\arcsec\,) 
in cyan and 843\,MHz contours
(at 5, 7, 10, 15, 20, 30, 40, and 50\,$\sigma$, where $\sigma \sim$\,5.57\,mJy; resolution\,$\sim$\,56\arcsec$\times$43\arcsec\,)
in blue. 
The symbols are same as for Fig. \ref{fig_NNdensity_extinction}.}  
\label{fig_radio} 
\end{figure*}

\begin{figure*}
\centering
\subfigure
{
\includegraphics[scale=0.5]{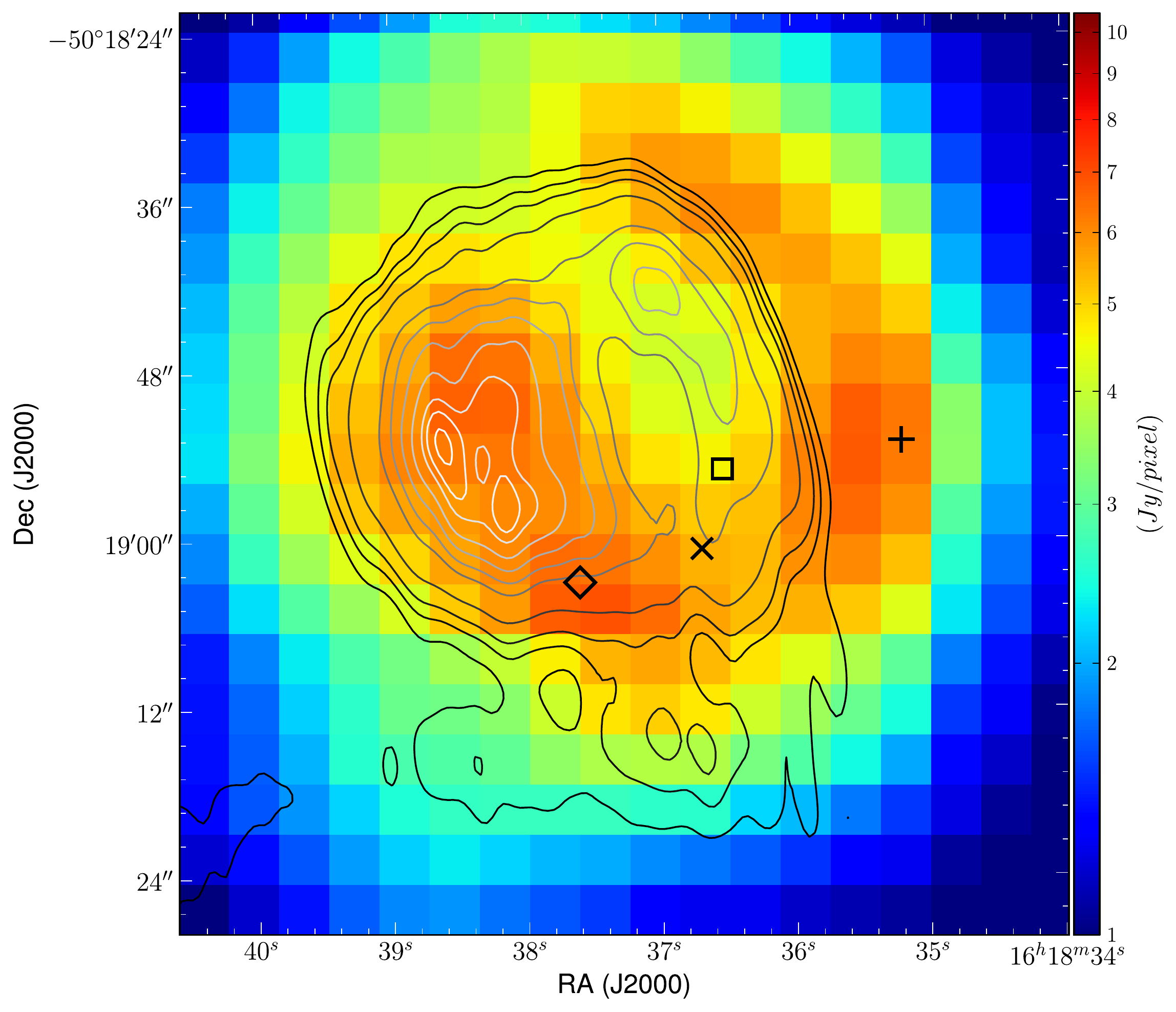}
}
\caption{The colourmap \emph{Herschel} 70\,\micron\, image with overlaid 1280\,MHz contours at 3, 4, 5, 7, 11, 13, 15, 
17, 19, 20, and 21\,$\sigma$ (where $\sigma \sim$\,0.4\,mJy). The contours are from the maximum resolution 
(7\arcsec$\times$2\arcsec) radio continuum image that could be constructed at 1280\,MHz. The symbols are same as 
Fig. \ref{fig_NNdensity_extinction}.}  
\label{fig_HighRes1280} 
\end{figure*}

Fig. \ref{fig_radio} shows a $Spitzer$ 8.0\,\micron\, image of the region with overlaid 843\,MHz contours from MGPS and 
1280\,MHz contours from GMRT. Rest of the objects marked are same as in Fig. \ref{fig_NNdensity_extinction}. 
The central core appears to be a compact \HII\, region, near whose 
peak lie the mm peak and the high-mass source. The 843\,MHz contours show extended emission in addition to the central compact 
region. 
The extended emission is only in the southern part and none in the northern part of the compact \HII\, region, indicating 
the presence of dense molecular cloud in the northern part which is not ionized to the same extent as the southern region.
The background 8.0\,\micron\, image also shows the diffuse nebular emission in the north.  
Using the {\sc aips} task {\sc jmfit}, the compact cores at both the frequencies were fit with a Gaussian model to determine 
the source sizes and the fluxes. The obtained results are given in Table \ref{table_Radio}. The beam-deconvolved source size 
for MGPS 843\,MHz is found to be much larger in area than that for 1280\,MHz. The larger integrated flux density for 843\,MHz 
could be because of this. Fig. \ref{fig_HighRes1280} shows the maximum resolution image of the region which could be 
obtained at 1280\,MHz ($\sim$7\arcsec$\times$2\arcsec), overlaid on the \emph{Herschel} 70\,\micron\, image. This contours show 
multiple peaks which is probably indicative of the clumpy nature of the ionized matter in the region. These peaks are mostly 
conincident with large 70\,\micron\, emission, which is to be expected as 70\,\micron\, also traces the thermal dust emission. 

We tried to estimate the physical parameters of the region, using the lower resolution images, as follows. 
The Lyman continuum luminosity (in photons s$^{-1}$) required to generate the observed flux density was determined using the 
following formula \citep[adapted from][see their Equations 1 and 3]{kurtz94} : 
\begin{equation}
S_{*} \geq \left(\frac{7.59\times10^{48}}{a(\nu, T_e)}\right)
           \left(\frac{S_{\nu}}{Jy}\right)
           \left(\frac{T_{e}}{K}\right)^{-0.5}
           \left(\frac{D}{kpc}\right)^{2}
           \left(\frac{\nu}{GHz}\right)^{0.1}
\end{equation}
where $S_{\nu}$ is the integrated flux density in Jy, $D$ is the distance in kiloparsec, $T_{e}$ is the electron temperature, 
$a(\nu, T_e)$ is the correction factor, and $\nu$ is the frequency in GHz at which the luminosity is to be calculated. 
The dynamical age of the \HII\, region ($t$) can be solved for by using the following equation from \citet{spitzer78} : 
\begin{equation}
R(t)=R_{s}\left(1+\frac{7c_{II}t}{4R_{s}}\right)^{4/7}
\end{equation}
where $R(t)$ is the radius of the H~{\sc ii} region at time $t$, $c_{II}$ is the speed of sound in \HII\, region 
\citep[11$\times$10$^{5}$ cm s$^{-1}$;][]{stahler05}, and $R_s$ is the Str\"omgren radius. The Str\"omgren radius ($R_s$, in 
cm) is given by \citep{stromgren39} : 
\begin{equation}
R_{s}=\left(\frac{3S_{*}}{4\pi n_{o}^{2}\beta_{2}}\right)^{1/3}
\end{equation}
where $n_{o}$ is the initial ambient density (in cm$^{-3}$), and $\beta_{2}$ is the total recombination coefficient to the first 
excited state of hydrogen. For our calculation, we assume a typical value of 10000\,K for $T_e$ \citep[which will imply a value 
of 0.99 for the correction factor \textquoteleft$a$\textquoteright\,; see Table\,6 of ][]{mezgerhenderson67}, and the corresponding 
$\beta_{2}$ of 2.6$\times$10$^{-13}$ cm$^{3}$ s$^{-1}$ \citep{stahler05}. For $n_o$, we use the value of 
4.8$\times$10$^4$\,cm$^{-3}$ from \citet{beltran06}. $R(t)$ is taken as the geometric mean of fitted Gaussian source sizes from 
Table \ref{table_Radio}. Using these formulae and the 1280\,MHz data, we calculated 
$S_*$ and $t$ as $\sim$\,10$^{47.41}$\,photons\,s$^{-1}$ and $\sim$\,0.3\,Myr, respectively. 
If we were to use the 843\,MHz flux density data point, then $S_*$ and $t$ come out to be $\sim$\,10$^{47.73}$\,photons\,s$^{-1}$ 
and $\sim$\,0.5\,Myr, respectively. 
Assuming ZAMS, a comparison of $\log S_*$ with the tabulated values from \citet{panagia73} shows that a spectral type of 
B0-O9.5 corresponds to this luminosity. Recent calibrations, like \citet{martins05}, also suggest a (luminosity class V) 
spectral type of $\sim$\,O9.5. 
So, it seems that the spectral type of the source ionizing the region (assuming a single source) has to be earlier than B0-O9.5  
for the ionization in the nebula to be sustained, since there can be absorption of ionizing photons by dust in the region which 
is often significant \citep{arthur04}. The dynamical age is approximately of the order of a few tenths of Myr. 

\begin{figure*}
\centering
\subfigure
{
\includegraphics[scale=0.5]{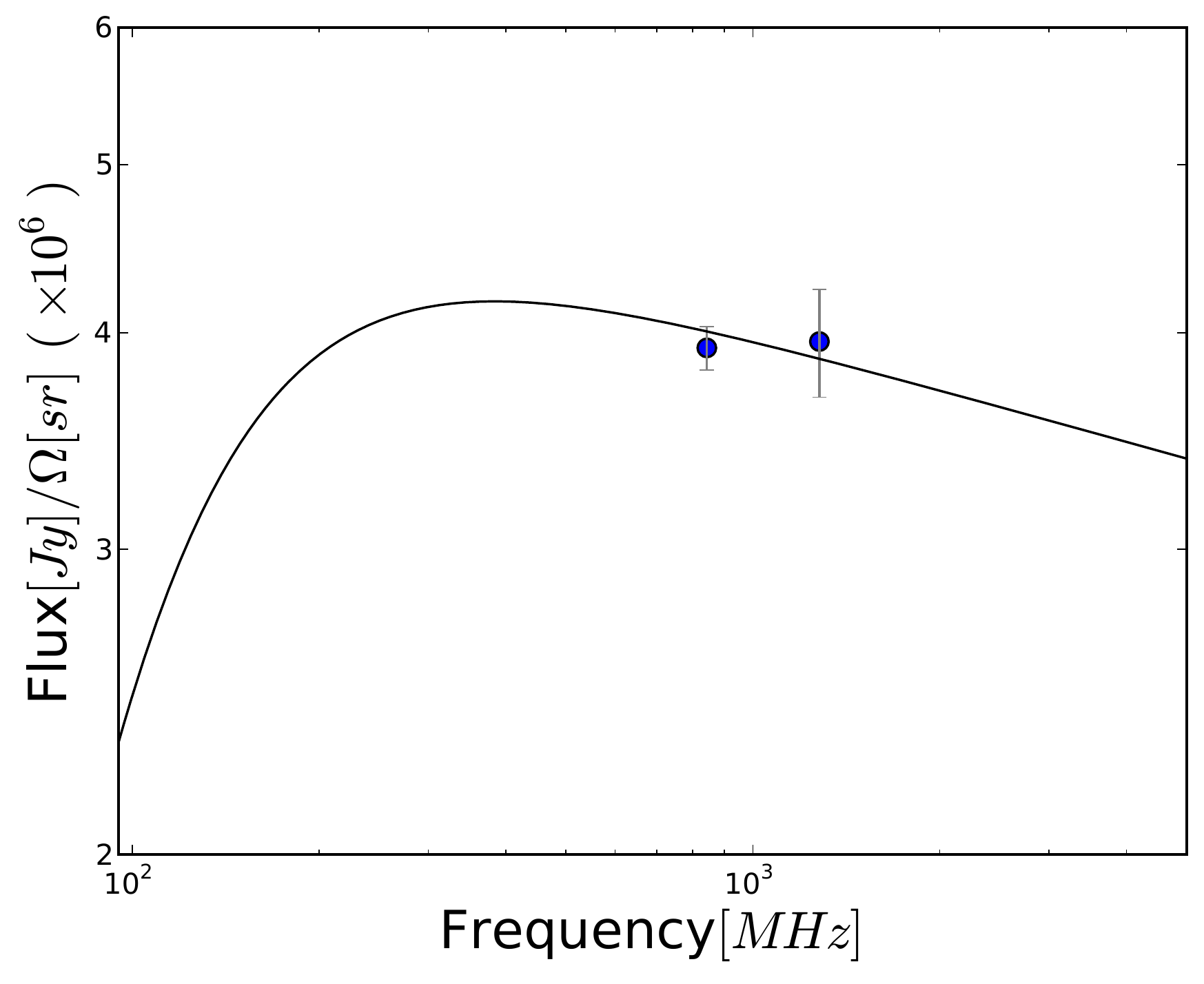}
}
\caption{The fitted free-free emission model for the \HII\, region. The data points at 843\,MHz and 1280\,MHz have been 
marked with solid circles.}  
\label{fig_EMfit} 
\end{figure*}

We also try to fit to our data the free-free emission model of \citet{mezgerhenderson67}, according to which 
\citep[adapted from ][]{mezgeretal67} : 

\begin{equation}
 S_{\nu}/\Omega=3.07\times10^{-2}T_{e}\nu^{2} (1-e^{-\tau(\nu)})
\end{equation}

\begin{equation}
 \tau(\nu)=1.643\times10^{5}a(\nu, T_e)T_{e}^{-1.35}\nu^{-2.1}n_{e}^{2}l
\end{equation}

where, $S_{\nu}$ is the integrated flux density (in Jy), $\nu$ is the frequency (in MHz), $n_{e}$ is the electron 
density in cm$^{-3}$, $l$ is the extent of the ionized region in pc, $\tau$ is the optical depth, and 
$\Omega$ is the solid angle subtended by the source (in steradians). $n_{e}^{2}l$ measures the optical depth in the medium 
(in cm$^{-6}$\,pc), and is called the emission measure. Taking $\Omega$ as 
($1.133\times\theta_{major}\times\theta_{minor}$), 
and the two data points, we fit the above equation using non-linear regression keeping $n_e^2 l$ - the emission measure - as a 
free parameter. The fit is shown in Fig. \ref{fig_EMfit}. The emission measure is thus determined to be 
$\sim$ 4.00\,$\pm$\,0.09\,$\times$\,10$^4$\,cm$^{-6}$\,pc. Since most of the central radio emission is confined within a circle 
of 120\arcsec\, diameter, we can assume it to be the extent of the \HII\, region ($\sim$\,2.1\,pc), and thus the electron 
density ($n_e$) 
turns out to be $\sim$\,138\,$cm^{-3}$. Further, using the formula from \citet[Equation\,A.5]{mezgeretal67}, we calculate the 
total mass of ionized hydrogen (M$_{\rm H\textsc{ii}}$) to be $\sim$\,16.4\,\msun\,. These low values of $n_e$ and 
M$_{\rm H\textsc{ii}}$, as well as the 
extent, would suggest that this region might be slightly more evolved than a compact \HII\, region \citep{kurtz02}.

\subsection{Central region} 
\label{section_CentralRegion} 

\begin{figure*}
\centering
\subfigure
{
\includegraphics[scale=0.8]{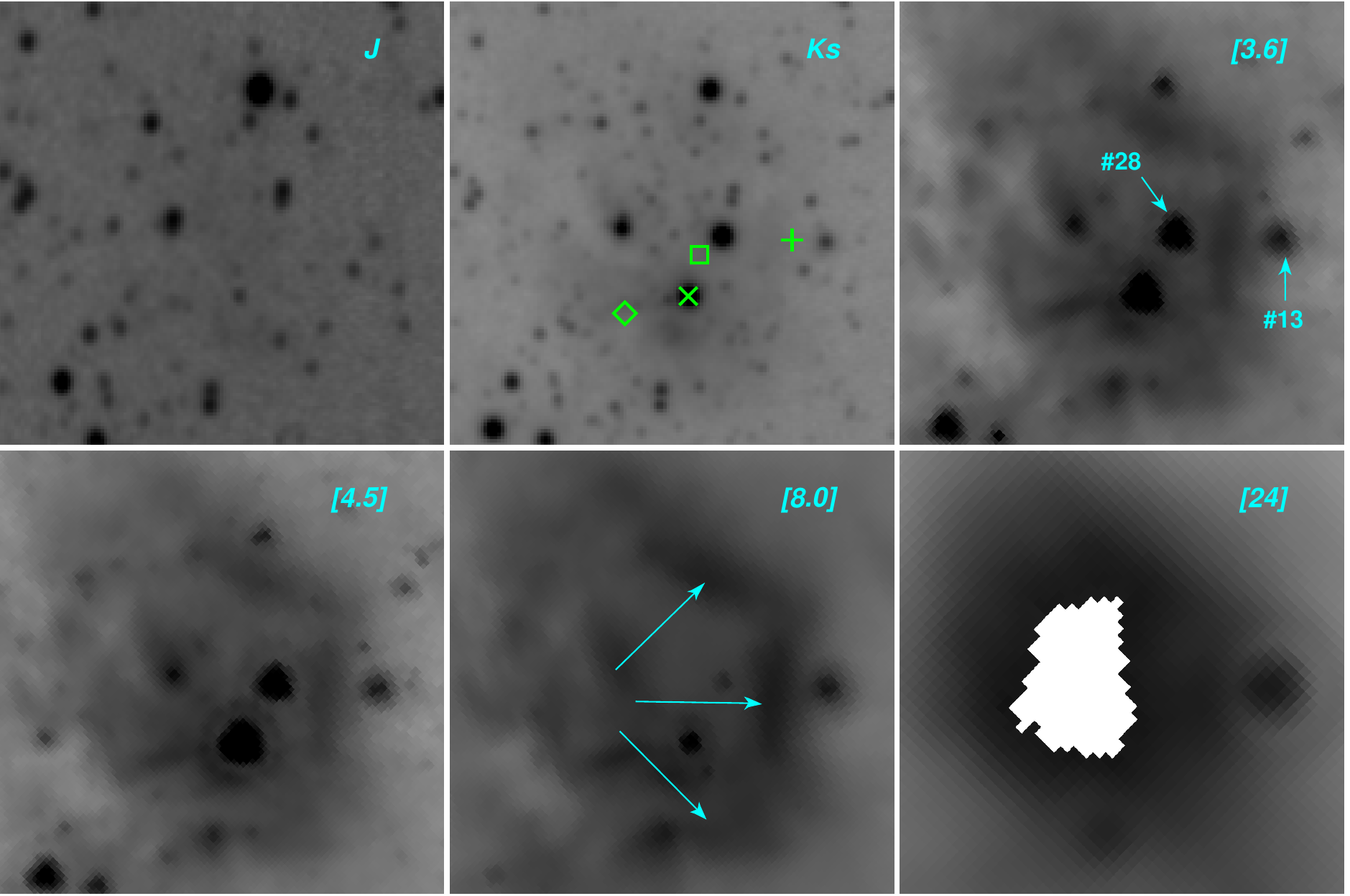}
}
\caption{1\arcmin\,$\times$\,1\arcmin\, central region of IRAS\,16148-5011. The images (left-to-right, top-to-bottom) are 
$J$, $K_s$, 3.6\,\micron\,, 4.5\,\micron\,, 8.0\,\micron\,, and 24\,\micron\,. North is up and east is to the left. The 
symbols on the $K_s$ band image are same as Fig. \ref{fig_NNdensity_extinction} (green cross is \#16 from Table \ref{table_SED}). 
The numbered sources on the 3.6\,\micron\, image are from Table \ref{table_SED}. The cyan arrows on the 8.0\,\micron\, image mark 
the \textquoteleft ring-like\textquoteright\, morphology. The white patch on the 24\,\micron\, image is the saturated region.}  
\label{fig_CentralRegion} 
\end{figure*}

Fig. \ref{fig_CentralRegion} shows the central 1\arcmin\,$\times$\,1\arcmin\, region of IRAS\,16148-5011 in the NIR $J$ and 
$K_s$ bands, IRAC 3.6\,\micron\,, 4.5\,\micron\,, and 8.0\,\micron\, bands, and the MIPS 24\,\micron\, band. 
The IRAC 3.6\,\micron\, and 8.0\,\micron\, bands, besides the continuum emission, also encompass the weak 
PAH emission feature at 3.3\,\micron\,, and strong PAH features at 7.7\,\micron\, and 8.6\,\micron\,. The 
4.5\,\micron\, band does not contain any PAH features, but does contain shocked molecular gas emission from 
$H_2(v=0-0)~S(9,10,11)$ and $CO(v=1-0)$. 24\,\micron\, emission is mainly the thermal continuum emission from the hot dust
\citep{watson08,churchwell09}.  

The $K_S$ band image has been marked with, among others, the high-mass source detected 
(green cross, \#16 in Table \ref{table_SED}) and the MSX peak from 
\citet[\textquoteleft IRAS\,16148-5011\,A\textquoteright]{molinari08} (green box). The marked high-mass source is of $\sim$ 
B2-B3 spectral type \citep[as per tabulated values from][see their Table\,1.1]{stahler05}, 
while the radio flux suggests an ionizing star of type $\geq$\,B0-O9.5. In 
addition, \citet{molinari08}, using SED modelling (assuming an embedded ZAMS source) and the MSX flux values, have estimated the 
spectral type of \textquoteleft IRAS\,16148-5011\,A\textquoteright\, source as O8. Therefore, it appears that there could be 
further embedded high-mass source(s) in this region, in addition to those seen in NIR and MIR, for the values to be consistent. 
It is possible that the cores seen by radio contours in Fig. \ref{fig_HighRes1280} could be hosting such high-mass source(s) 
and contributing to the radio luminosity. 

The other sources associated with this central part of the nebula have been marked on the 3.6\,\micron\, image (\#13 and \#28 
from Table \ref{table_SED}, Section \ref{section_SED}). The source \#13 appears to be a highly embedded intermediate-mass 
young source ($\sim$\,6.17\,\msun\,, $\sim$\,a few 10$^4$\,yr; see Section \ref{section_SED}), even visible at 24\,\micron. 
Source \#28 appears to be a much older and evolved 
source ($\sim$\,8.48\,\msun\,, $\sim$\,a few 10$^6$\,yr), possibly shrouded in the surrounding nebulosity (interstellar 
extinction of A$_V \sim$\,18.38\,mag), leading to a lack of detection in the $J$ band (though it is the second brightest source 
after the 9.6\,\msun\, - \#16 from Table \ref{table_SED} - source in the central region in $K_s$ band). The radio continuum 
emission contribution from this source will be much lower, hardly making a difference even when taken in combination with the 
emission from the high-mass source, and thus will not affect the inferences regarding radio emission above. 

For \HII\, regions, the PAH emissions serve as useful diagnostic tools as they are characteristic of photo-dissociation regions 
(PDRs). As a high-mass star ionizes its natal medium, 
the UV radiation destroys the PAH in its surroundings. However, at the boundary of the \HII\, region produced, the UV intensity 
falls off, and a PDR is formed where the PAH's are merely highly excited, leading to strong emission \citep{povich07}. 
The PDR is supposed to be the transition region between the ionized and neutral matter. This 
usually results in a \textquoteleft bubble\textquoteright\, morphology, where \HII\, regions surrounded by ring-like PAH 
emission are seen \citep{churchwell06}. A similar morphology can be seen in the 3.6\,\micron\, and the 8.0\,\micron\, image 
in Fig. \ref{fig_CentralRegion}. A \textquoteleft semi-ring\textquoteright\, in the western half of the image is seen (marked by 
cyan arrows on the 8.0\,\micron\, image). We note that this feature is also faintly seen in the 4.5\,\micron\, image, though this 
could just be continuum emission. 
That this ring-like feature does not appear symmetric (i.e. there is no clear eastern \textquoteleft semi-ring\textquoteright\,, 
though faint indications are seen) could possibly be due to eastern denser and/or non-homogeneous molecular cloud, or projection 
effects. \citet{volk91}, on the basis of low-resolution spectra, had classified 
IRAS\,16148-5011 as a PAH source. Bulk of the radio emission (see Fig. \ref{fig_radio}), as well as the 
24\,\micron\, emission from hot dust, is confined within this ring-like feature as expected \citep{watson08}. Some of the 
emission from this ring-like structure could also be due to swept-up material by the expanding ionization front of the \HII\, 
region.

\subsection{\textit{Herschel} Results}
\label{section_HerschelResults}

Thermal emission from cold dust lies in the FIR wavelength range, and thus its analysis can be used to obtain the physical 
parameters like dust temperature and column density of a region 
\citep{launhardt13, battersby11}. This was 
carried out by the SED modelling of the thermal dust emission, whose Rayleigh-Jeans regime is covered by the \textit{Herschel} 
FIR bands (160--500\,\micron), using the following order of steps.  

First of all, the surface brightness unit for all the images was converted to Jy\,pixel$^{-1}$. Since the PACS image is already in 
Jy\,pixel$^{-1}$, this step was only required for the SPIRE images (whose units are in MJy\,sr$^{-1}$), and was carried out using 
the pixel scales for the respective SPIRE bands. 
Next, the 160-350\,\micron\, images were convolved to the resolution of the 500\,\micron\, image ($\sim$\,36\arcsec\,, lowest among 
all images) 
using the convolution kernels of \citet{aniano11}, and regridded to a pixel scale of 14\,\arcsec\, (same as 500\,\micron\, image). 
Using these final reworked images with same resolution and pixel scale, a background flux level, $I_{bg}$, was determined 
from a \textquotedblleft smooth\textquotedblright\, (i.e. no abrupt clumpy regions) and relatively 
\textquotedblleft dark\textquotedblright\, patch of the sky. 
The distribution of individual pixel values in the dark patch of the sky, for each of the bands, was fitted 
with a Gaussian iteratively, rejecting the pixel values 
outside $\pm$\,2$\sigma$ in each iteration, till the fit converged. The same patch of the sky was used for each band. The 
background flux level, $I_{bg}$, was thus determined as 0.29, 2.67, 1.27, and 0.45 Jy\,pixel$^{-1}$ for the 160\,\micron, 
250\,\micron, 350\,\micron, and 500\,\micron\, images, respectively. We note that the 70\,\micron\, image, though available, 
was not used here as the optically thin assumption might not hold true at this wavelength \citep{lombardi14}. 

\begin{figure*}
\centering
\subfigure
{
\includegraphics[scale=0.5]{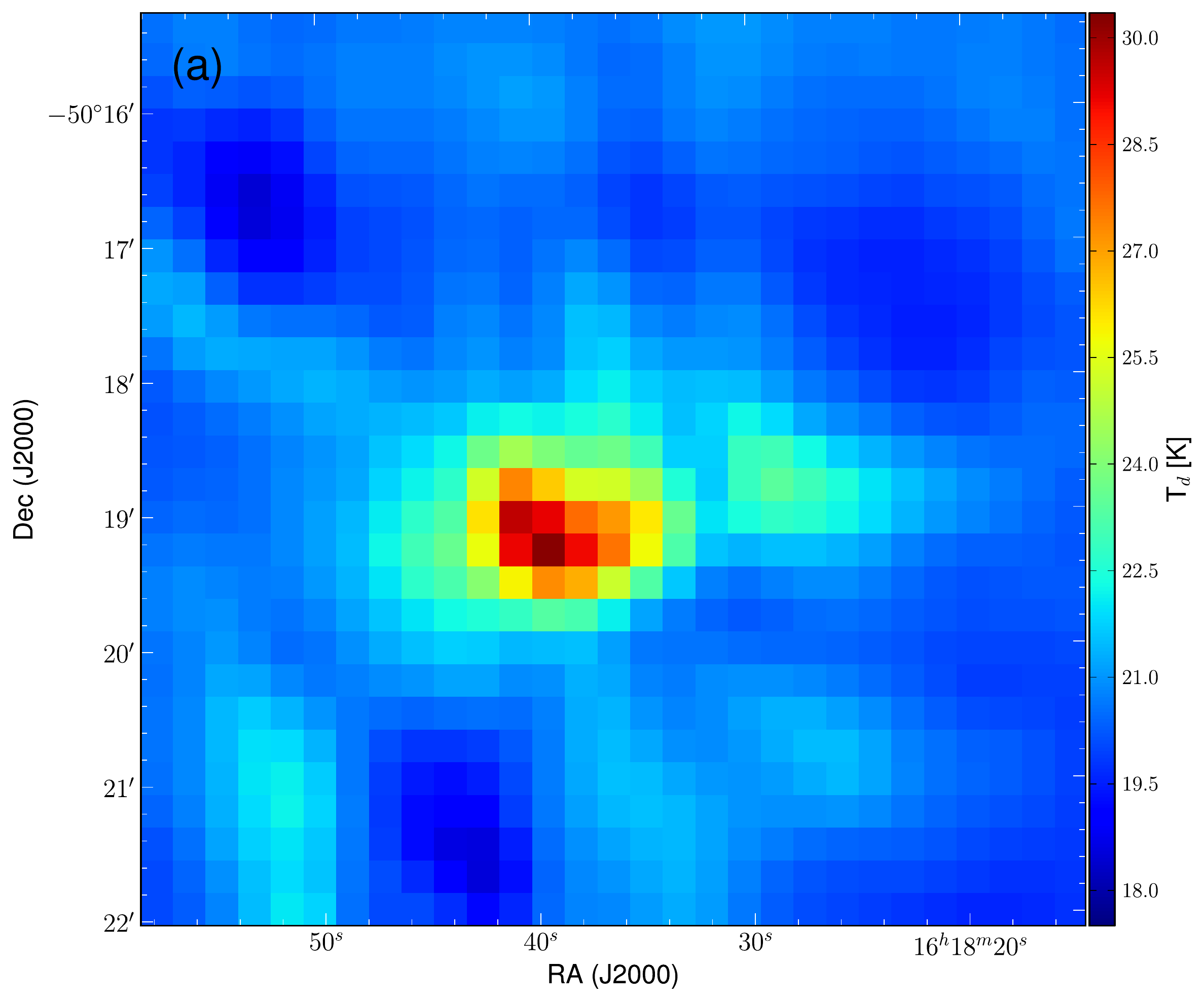}
\label{fig_HerschelMaps_Temperature}
}
\subfigure
{
\includegraphics[scale=0.5]{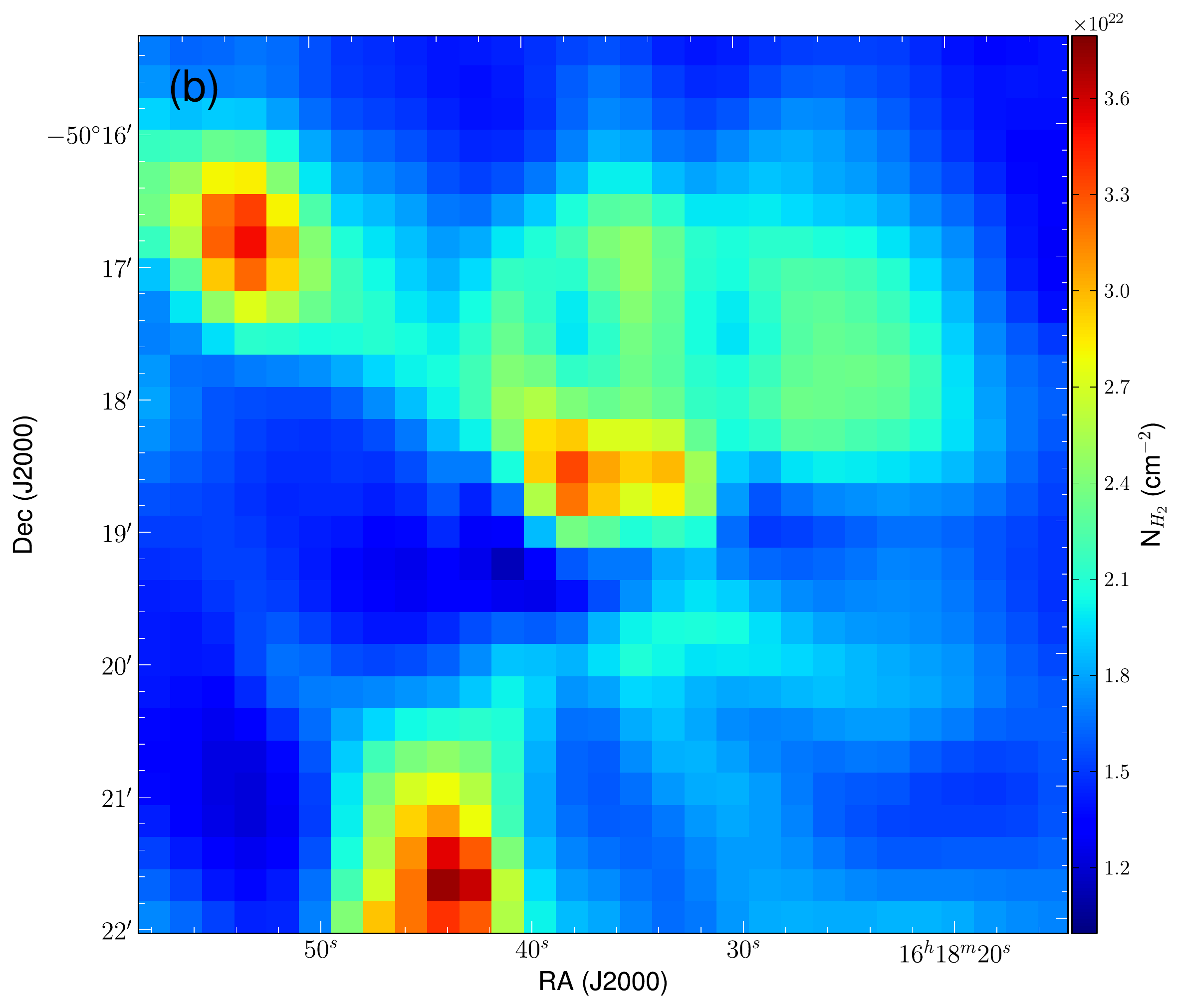}
\label{fig_HerschelMaps_ColumnDensity} 
}
\caption{(a) Dust temperature map, and (b) column density map of the region around IRAS\,16148-5011, derived using the SED fitting 
to the thermal dust emission.}  
\label{fig_HerschelMaps} 
\end{figure*}

Modified blackbody fitting was subsequently carried out on a pixel-by-pixel basis using the following formulation 
\citep{battersby11, sadavoy12, nielbock12, launhardt13} :  
\begin{equation}
\label{equation_ModifiedBlackbody} 
S_{\nu}(\nu) - I_{bg}(\nu) = B_{\nu}(\nu, T_d) \Omega (1-e^{-\tau (\nu)}),  
\end{equation}
with  
\begin{equation}
\tau (\nu) = \mu_{H_2} m_H \kappa_{\nu} N(H_2),  
\end{equation}
where, $\nu$ is the frequency, $S_{\nu}(\nu)$ is the observed flux density, $I_{bg}(\nu)$ is the background flux in that 
particular band 
(estimated above), $B_{\nu}(\nu, T_d)$ is the Planck's function, $T_d$ is the dust temperature, $\Omega$ is the solid angle 
(in steradians) from where the flux is obtained (just the solid angle subtended by a 14\arcsec\,$\times$14\arcsec\, pixel here), 
$\tau(\nu)$ is the optical depth, $\mu_{H_2}$ is the mean molecular weight (adopted as 2.8 here), $m_H$ is the mass of 
hydrogen, $\kappa_{\nu}$ is the dust opacity, and $N(H_2)$ is the column density. For opacity, we adopt a functional form 
of $\kappa_{\nu} = 0.1$\,$(\nu/1000$\,GHz$)^{\beta}$ cm$^2$\,g$^{-1}$, with $\beta=2$ 
(see \citealt{andre10}, \citealt{beckwith90}, \citealt{hildebrand83}).   
For each pixel, Equation \ref{equation_ModifiedBlackbody} was fit using the 4 data points, keeping $T_d$ and $N(H_2)$ as free 
parameters. Pixels for which the fit did not converge, or the error was larger than 10\%, had their values taken as the median 
of 8 immediate-neighbour pixels. The final obtained temperature and column density maps of the wider region (to clearly discern 
the morphological features) surrounding IRAS\,16148-5011 are shown in Fig. \ref{fig_HerschelMaps}. 

From the temperature (Fig. \ref{fig_HerschelMaps_Temperature}) and column density (Fig. \ref{fig_HerschelMaps_ColumnDensity}) 
maps, we obtain the peak values as $\sim$\,30\,K and 3.3$\times$10$^{22}$\,cm$^{-2}$, respectively, for IRAS\,16148-5011.  
\citet{fontani05} provide a temperature of 38\,K (using greybody fit with 60\,\micron\,, 100\,\micron\,, and 1.2\,mm data), and 
a (beam-averaged) column density of 3$\times$10$^{23}$\,cm$^{-2}$ (using C$^{17}$O molecular line observations) for 
IRAS\,16148-5011. Here our peak 
temperature is about 20\% lower, and the column density estimate an order of magnitude smaller. The difference in temperature 
seems to be due to their fitting limitations owing to sparse data points, as well as the fact that the 60\,\micron\, emission 
might not be 
optically thin. It should be noted, however, that the dust temperature distribution from \citet{fontani05} (for their analysed 
catalogue of IRAS sources) peaks at $\sim$\,30\,K. Column density disparity can probably be explained by the dependence of 
\citet{fontani05} on the molecular abundance of the rare C$^{17}$O, whose value can have wide variations \citep{redman02,walsh10}. 

The column density map (Fig. \ref{fig_HerschelMaps_ColumnDensity}) displays three peaks - the central IRAS\,16148-5011 object, a 
peak to its north-east, and a peak to its south. These north-east and south peaks appear to be associated with cold clumps, as is 
evident from the temperature map which shows T$_d < 20$\,K at their positions. The column density map shows that all the peaks 
appear to be connected by weak filamententary features, similar to previous results for myriad regions 
\citep[see][and references therein]{andre13}.  
For the central IRAS\,16148-5011 object, we estimated the associated clump dimensions, using the 
\textquotedblleft clumpfind\textquotedblright\, software \citep{williams94}, to be 
$\sim$\,62\arcsec$\times$46\arcsec\, (i.e. 1.1\,pc$\times$0.8\,pc at 3.6\,kpc). Due to 
the low resolution of the column density image generated here, it does not appear possible to resolve further sub-clumps. 
The mass of the clump can be estimated by :
\begin{eqnarray}
M_{clump} &=& \mu_{H_2}~m_{H}~\Sigma ~ N(H_2)~Area_{pixel} \\
          &=& \mu_{H_2}~m_{H}~Area_{pixel}~\Sigma ~ N(H_2)
\end{eqnarray}
i.e. calculating the mass in each pixel and then summing over all the pixels which constitute the clump. Using the 
$\Sigma ~ N(H_2)$ returned by the \textquotedblleft clumpfind\textquotedblright\, software, the total clump mass was calculated 
to be $\sim$\,1023\,\msun. 

Another noticeable feature in this map is that the immediate northern vicinity of the central region exhibits higher column 
density than the immediate southern portion. This suggests dense nebula in the northern part, affirming the inference also drawn
from extinction and cluster analysis in Section \ref{section_ClusterAndExtinction} (also see Fig. \ref{fig_NNdensity_extinction}).
This is also consonant with the fact that the dense nebular emission is in the northern part, with no extended radio emission 
seen there (unlike in the southern part). To get an estimate of the visual extinction 
in this northern part, we use the relation $\langle N(H_2)/A_V \rangle$=0.94$\times$10$^{21}$\,molecules\,cm$^{-1}$\,mag$^{-1}$ 
\citep[adapted from][assuming a total-to-selective extinction ratio $R_V$=3.1, and that the gas is in molecular form]{bohlin78}. 
Now, typical N(H$_2$) seen here is $\gtrsim$\,2$\times$10$^{22}$\,cm$^{-3}$, which implies an A$_V >20$\,mag. This high value of 
A$_V$ is probably the reason why very few YSOs are associated with this northern nebular emission 
(see Fig. \ref{fig_NNdensity_extinction}). If the value of R$_V$ were to be larger, as has been conjectured for dense 
environments, then it will lead to an increase in A$_V$.

\section{Discussion} 
\label{section_Discussion} 

The IRAS\,16148-5011 region appears to be hosting an infrared cluster containing high-mass star(s) embedded in the nebula, 
and could serve as a future template to study the high-mass star formation process. 
Age estimates for any cluster are usually plagued with uncertainties \citep[see][for a discussion]{lada03}, and various proxies 
are often used to ascertain the evolutionary stage of a cluster. One such is the ratio of Class\,II to Class\,I sources 
(as Class\,II sources are older) \citep{schmeja05, beerer10, gutermuth09}. We have detected 133 Class\,II and 7 Class\,I sources 
in this region. However, here it seems that the large ratio (19) could be because Class\,I sources are deeply 
embedded and their detection is affected by large nebular extinction, though varying ratio values have been estimated for other 
clusters \citep{gutermuth09}. 

The two brightest (NIR) sources in the central region appear to be high-mass (see Section \ref{section_CentralRegion}) with 
indications of futher embedded high-mass sources. Also in literature, \citet{beltran06} report two clumps within 90\arcsec\, 
of the IRAS\,16148-5011 IRAS catalogue position - \textquoteleft 16148-5011\,Clump\,2\textquoteright\, (206\,\msun) and 
\textquoteleft 16148-5011\,Clump\,3\textquoteright\, (42\,\msun) - detected using 1.2\,mm emission and a dust temperature of 
30\,K. While \textquoteleft Clump\,2\textquoteright\,, is almost coincident with the 
\textquoteleft 16148-5011\,MM\,1\textquoteright\, peak from \citet{molinari08} (marked with a diamond symbol on 
Fig. \ref{fig_CentralRegion}), \textquoteleft Clump\,3\textquoteright\, coordinates are same as the IRAS\,16148-5011's IRAS 
catalogue coordinates (green plus in Fig. \ref{fig_CentralRegion}). The presence of these massive clumps could also point towards 
ongoing high-mass star formation in the region. 

Though this part of the sky also harbours other IRAS sources \citep{karnik01}, IRAS\,16148-5011 does not appear to be a part of 
any larger star-forming complex as such, but it seems to be connected to other clumps via filaments \citep[similar to the 
hub-filament morphology of][]{myers09}. 
The high-mass stellar cluster formation in the associated molecular cloud appears to be spontaneous. However, along the rather 
sharp southern boundary of the molecular cloud, the results from cluster analysis show a stellar sub-cluster forming 
(Fig. \ref{fig_NNdensity_extinction}), which could partly be due to the triggering by the expanding ionization front of the \HII\, 
region.

\section{Conclusions} 
\label{section_conclusion}

The main conclusions of this paper, resulting from a multiwavelength study involving NIR, MIR, FIR, and radio continuum data, 
are as follows : 
\begin{enumerate}
\item 7 Class\,I and 133 Class\,II YSOs are identified using a combination of NIR and MIR data. 
A 9.6\,\msun\, high-mass source is found to be associated with the central nebula.

\item Low-frequency radio emission reveals a compact \HII\, region, with extended emission in the southern part. Lyman continuum 
photon luminosity calculation gives B0-O9.5 as the lower limit for the spectral type of the ionizing source (assuming single 
source). The dynamical age of the \HII\, region is in 0.3-0.5\,Myr range. 
SED modelling of the free-free emission yields an electron density of 138\,cm$^{-3}$. The mass of the ionized hydrogen is 
calculated to be $\sim$\,16.4\,\msun. The high-resolution 7\arcsec$\times$2\arcsec\, contour map shows clumpy ionized gas in the 
region. 

\item The central nebular region shows a ring-like PAH emission feature near the borders of the compact \HII\, region, tracing 
the PDR. There appear to be three NIR- and MIR-visible central sources, of masses $\sim$ 6.2\,\msun, 9.60\,\msun, and 
8.5\,\msun. Based on the incongruency between the total radio flux from these sources and the flux obtained from radio 
observations, and literature estimates of an early type star using MIR SED fitting, it is possible that there could be   
high-mass embedded source(s) present in this region. 

\item Dust temperature and column density maps are obtained using SED modelling of the thermal dust emission. The peak 
temperature and column density values are 30\,K and 3.3$\times$10$^{22}$\,cm$^{-2}$, respectively, for 
IRAS\,16148-5011. The column density map reveals that the immediate northern vicinity of IRAS\,16148-5011, which contains 
the nebular emission seen prominently at MIR wavelengths, has a large extinction of A$_V >$\,20\,mag. This map also shows 
that weak filamententary structures join IRAS\,16148-5011 to nearby cold clumps. 
The size and mass of the clump associated with IRAS\,16148-5011 is estimated to be $\sim$\,1.1\,pc$\times$0.8\,pc (at 3.6\,kpc) 
and $\sim$\,1023\,\msun, respectively. 
\end{enumerate}

Future observations of individual objects in spectral lines, deeper IR data to get a full stellar census upto below brown-dwarf 
limit, further molecular line observations which probe high column densities will help put the star formation scenario on a firm 
footing and help study high-mass star formation.

\section*{Acknowledgments}

We thank the anonymous referee for a thorough reading of the manuscript, and for the useful comments and suggestions which 
helped improve its scientific content. 
The authors thank the staff of IRSF in South Africa, a joint partnership between S.A.A.O and Nagoya University of Japan; 
and GMRT managed by National Center for Radio Astrophysics of the Tata Institute of Fundamental Research (TIFR) for their 
assistance and support during observations. D.K.O was supported by the National Astronomical Observatory of Japan (NAOJ), 
Mitaka, through a fellowship, during which a part of this work was done.

\begin{landscape}
\begin{table}
\centering 
\caption{YSOs identified using NIR and MIR data}
\label{table_YSOs} 
\begin{tabular}{@{}cccccccccc}
\hline 
RA & Dec. & $J$ & $H$ & $K_s$ & $[3.6]$ & $[4.5]$ & $[5.8]$ & $[8.0]$ & YSO  \\
(J2000) & (J2000) & (mag) & (mag) & (mag) & (mag) & (mag) & (mag) & (mag) & Classification \\
\hline 
 244.575165   &  -50.365025    &    17.861   $\pm$  0.130   &  16.631   $\pm$  0.001   &  15.509   $\pm$  0.047   &  14.311   $\pm$  0.145  &   14.132   $\pm$  0.203   &    ---   &     ---   & Class2  \\ 
 244.575241   &  -50.367393    &    14.580   $\pm$  0.021   &  13.876   $\pm$  0.010   &  13.643   $\pm$  0.032   &  12.983   $\pm$  0.099  &   12.872   $\pm$  0.122   &    ---   &     ---   & Class2  \\
 244.576645   &  -50.293938    &    16.137   $\pm$  0.016   &  14.443   $\pm$  0.010   &  13.570   $\pm$  0.020   &  12.780   $\pm$  0.102  &   12.372   $\pm$  0.132   &    ---   &     ---   & Class2  \\ 
 244.577545   &  -50.314171    &    15.557   $\pm$  0.010   &  14.358   $\pm$  0.010   &  13.818   $\pm$  0.028   &  13.001   $\pm$  0.108  &   12.820   $\pm$  0.143   &    ---   &     ---   & Class2  \\ 
 244.579193   &  -50.321396    &    17.491   $\pm$  0.045   &  15.921   $\pm$  0.020   &  14.881   $\pm$  0.038   &  13.412   $\pm$  0.161  &   13.265   $\pm$  0.132   &    ---   &     ---   & Class2  \\  
\hline
\end{tabular}

\begin{footnotesize}
Table~\ref{table_YSOs} is available in its entirety in a machine-readable form in the online journal. 
A portion is shown here for guidance regarding its form and content.
\end{footnotesize}
\end{table}
\end{landscape}

\begin{landscape}
\begin{table}
\centering 
\caption{Main parameters from SED analysis}
\label{table_SED} 
\begin{tabular}{@{}lccccccccccc}
\hline 
S.No. & RA       & Dec.   & $\log\,t_*$ & Mass           &  $\log\,M_{disk}$ & $\log\,\dot{M}_{disk}$ &  $\log\,M_{env}$   &  $\log\,T_*$   &  $\log\,L_{total}$  &  $A_V$  & $\chi_{min}^2$ \\
      &  (J2000) & (J2000) & ($yr$)       & ($M_{\odot}$) &  ($M_{\odot}$)   & ($M_{\odot}~yr^{-1}$) &  ($M_{\odot}$)    &  (\emph{K})   &  ($L_{\odot}$)     &  (\emph{mag})  &  (per data point) \\ 
\hline 
\multicolumn{12}{c}{Class\,I sources} \\
\hline
 1         &  244.591080     &   -50.285355     &   6.33      $\pm$   0.74      &   2.42      $\pm$   0.79      &   -3.38     $\pm$   1.64      &   -9.29     $\pm$   1.73      &   -4.43     $\pm$   2.73      &   3.85      $\pm$   0.15      &   1.41      $\pm$   0.30      &   4.81      $\pm$   2.35      &   0.29      \\      
 2         &  244.603455     &   -50.292828     &   4.71      $\pm$   0.40      &   4.20      $\pm$   1.42      &   -1.39     $\pm$   0.70      &   -6.70     $\pm$   0.99      &   0.68      $\pm$   1.06      &   3.66      $\pm$   0.07      &   2.11      $\pm$   0.28      &   15.95     $\pm$   4.98      &   0.01      \\      
 3         &  244.671539     &   -50.338348     &   5.30      $\pm$   1.02      &   2.58      $\pm$   1.61      &   -2.90     $\pm$   1.82      &   -8.08     $\pm$   2.02      &   -2.70     $\pm$   2.90      &   3.73      $\pm$   0.20      &   1.56      $\pm$   0.56      &   7.45      $\pm$   4.16      &   0.10      \\      
\hline
\multicolumn{12}{c}{Class\,II sources} \\
\hline
 4         &  244.584152     &   -50.353664     &   4.65      $\pm$   0.37      &   6.41      $\pm$   0.54      &   -2.12     $\pm$   0.75      &   -6.94     $\pm$   0.80      &   1.65      $\pm$   0.40      &   3.66      $\pm$   0.01      &   2.49      $\pm$   0.06      &   12.17     $\pm$   5.81      &   9.90      \\      
 5         &  244.599808     &   -50.286560     &   4.94      $\pm$   0.37      &   3.80      $\pm$   1.81      &   -1.99     $\pm$   0.84      &   -7.55     $\pm$   1.30      &   0.54      $\pm$   0.66      &   3.66      $\pm$   0.05      &   1.93      $\pm$   0.41      &   11.51     $\pm$   5.27      &   1.93      \\      
 6         &  244.606796     &   -50.307327     &   5.57      $\pm$   0.87      &   7.50      $\pm$   3.12      &   -4.67     $\pm$   2.72      &   -9.18     $\pm$   2.08      &   -3.10     $\pm$   4.21      &   4.02      $\pm$   0.35      &   2.95      $\pm$   0.68      &   13.72     $\pm$   5.07      &   5.34      \\      
 7         &  244.609207     &   -50.351631     &   4.97      $\pm$   0.27      &   4.68      $\pm$   1.40      &   -2.11     $\pm$   0.93      &   -7.66     $\pm$   1.31      &   1.24      $\pm$   0.53      &   3.67      $\pm$   0.05      &   2.12      $\pm$   0.32      &   17.37     $\pm$   3.72      &   10.85     \\      
 8         &  244.610397     &   -50.279411     &   5.14      $\pm$   0.76      &   2.99      $\pm$   1.59      &   -2.25     $\pm$   0.95      &   -7.76     $\pm$   1.28      &   -0.70     $\pm$   2.22      &   3.70      $\pm$   0.15      &   1.73      $\pm$   0.36      &   13.43     $\pm$   5.74      &   0.29      \\      
 9         &  244.612305     &   -50.365067     &   4.84      $\pm$   0.45      &   5.60      $\pm$   2.59      &   -2.04     $\pm$   1.05      &   -7.61     $\pm$   1.55      &   1.09      $\pm$   0.87      &   3.70      $\pm$   0.07      &   2.31      $\pm$   0.66      &   4.51      $\pm$   2.81      &   0.34      \\      
10         &  244.614349     &   -50.327496     &   6.28      $\pm$   0.66      &   2.50      $\pm$   0.83      &   -3.40     $\pm$   1.71      &   -9.21     $\pm$   1.89      &   -4.37     $\pm$   2.56      &   3.85      $\pm$   0.17      &   1.44      $\pm$   0.35      &   6.20      $\pm$   2.30      &   0.13      \\      
11         &  244.633514     &   -50.326122     &   5.73      $\pm$   0.70      &   2.58      $\pm$   0.99      &   -3.20     $\pm$   1.77      &   -8.84     $\pm$   1.80      &   -2.83     $\pm$   2.80      &   3.77      $\pm$   0.19      &   1.54      $\pm$   0.35      &   8.83      $\pm$   4.50      &   0.28      \\      
12         &  244.639954     &   -50.283981     &   5.10      $\pm$   0.07      &   4.80      $\pm$   0.22      &   -2.36     $\pm$   0.66      &   -7.50     $\pm$   0.20      &   1.14      $\pm$   0.56      &   3.66      $\pm$   0.01      &   2.02      $\pm$   0.07      &   9.75      $\pm$   5.38      &   2.16      \\      
13$^{a,b}$ &  244.644714     &   -50.314804     &   4.18      $\pm$   1.01      &   6.17      $\pm$   1.74      &                        -      &                        -      &   -0.18     $\pm$   1.92      &   3.76      $\pm$   0.26      &   2.82      $\pm$   0.31      &   6.18      $\pm$   3.86      &   0.03      \\      
14         &  244.649155     &   -50.346909     &   5.65      $\pm$   0.90      &   2.44      $\pm$   1.12      &   -2.92     $\pm$   1.44      &   -8.40     $\pm$   1.68      &   -2.87     $\pm$   3.05      &   3.79      $\pm$   0.20      &   1.58      $\pm$   0.35      &   7.63      $\pm$   4.29      &   0.03      \\      
15         &  244.651047     &   -50.328751     &   6.29      $\pm$   0.68      &   3.46      $\pm$   0.86      &   -3.43     $\pm$   1.48      &   -8.96     $\pm$   1.60      &   -4.83     $\pm$   2.75      &   3.99      $\pm$   0.18      &   2.01      $\pm$   0.27      &   6.70      $\pm$   2.41      &   0.13      \\      
16$^{a,c}$   &  244.652878     &   -50.316849     &   5.44      $\pm$   0.61      &   9.60      $\pm$   1.44      &   -2.15     $\pm$   1.24      &   -7.00     $\pm$   1.35      &    0.00     $\pm$   3.42      &   4.29      $\pm$   0.20      &   3.69      $\pm$   0.16      &   15.97     $\pm$   4.36      &   0.89      \\      
17         &  244.663803     &   -50.352959     &   6.02      $\pm$   0.78      &   2.93      $\pm$   0.82      &   -3.23     $\pm$   1.35      &   -8.92     $\pm$   1.53      &   -3.75     $\pm$   3.08      &   3.86      $\pm$   0.19      &   1.74      $\pm$   0.24      &   4.75      $\pm$   2.79      &   0.07      \\      
18         &  244.671738     &   -50.278400     &   5.30      $\pm$   0.28      &   3.02      $\pm$   1.19      &   -2.97     $\pm$   1.18      &   -8.96     $\pm$   1.72      &   -0.74     $\pm$   0.91      &   3.67      $\pm$   0.05      &   1.62      $\pm$   0.27      &   10.13     $\pm$   3.37      &   3.23      \\      
19         &  244.679535     &   -50.293499     &   4.97      $\pm$   0.36      &   4.10      $\pm$   1.47      &   -2.07     $\pm$   0.86      &   -7.70     $\pm$   1.19      &   0.45      $\pm$   0.86      &   3.67      $\pm$   0.06      &   2.01      $\pm$   0.30      &   9.06      $\pm$   5.71      &   2.25      \\      
20         &  244.685440     &   -50.357010     &   6.29      $\pm$   0.72      &   3.69      $\pm$   0.99      &   -3.83     $\pm$   1.67      &   -9.31     $\pm$   1.77      &   -5.06     $\pm$   2.77      &   4.03      $\pm$   0.19      &   2.14      $\pm$   0.32      &   9.15      $\pm$   3.12      &   0.17      \\      
21         &  244.689316     &   -50.343143     &   4.99      $\pm$   0.39      &   3.85      $\pm$   1.38      &   -2.13     $\pm$   0.83      &   -7.74     $\pm$   1.20      &   0.28      $\pm$   1.05      &   3.67      $\pm$   0.06      &   1.94      $\pm$   0.26      &   7.45      $\pm$   5.58      &   2.23      \\      
22         &  244.689453     &   -50.366440     &   4.90      $\pm$   0.48      &   4.06      $\pm$   1.33      &   -1.97     $\pm$   0.87      &   -7.40     $\pm$   1.17      &   0.37      $\pm$   1.39      &   3.67      $\pm$   0.09      &   2.04      $\pm$   0.26      &   13.02     $\pm$   5.40      &   0.17      \\      
23         &  244.693146     &   -50.279778     &   5.50      $\pm$   0.67      &   2.95      $\pm$   0.96      &   -2.39     $\pm$   0.94      &   -8.01     $\pm$   1.08      &   -1.47     $\pm$   2.53      &   3.73      $\pm$   0.16      &   1.68      $\pm$   0.19      &   7.19      $\pm$   3.64      &   4.84      \\      
24         &  244.704742     &   -50.351349     &   4.78      $\pm$   0.39      &   5.70      $\pm$   0.86      &   -2.26     $\pm$   0.30      &   -7.61     $\pm$   0.94      &   1.65      $\pm$   0.30      &   3.66      $\pm$   0.02      &   2.37      $\pm$   0.09      &   17.61     $\pm$   2.63      &   14.26     \\      
25         &  244.708603     &   -50.286385     &   4.89      $\pm$   0.35      &   4.59      $\pm$   1.48      &   -1.92     $\pm$   0.84      &   -7.52     $\pm$   1.14      &   0.89      $\pm$   0.71      &   3.67      $\pm$   0.06      &   2.14      $\pm$   0.34      &   12.96     $\pm$   5.63      &   3.55      \\      
26         &  244.683786     &   -50.363702     &   4.75      $\pm$   0.39      &   6.24      $\pm$   2.25      &   -1.56     $\pm$   0.82      &   -6.91     $\pm$   1.09      &   1.25      $\pm$   0.72      &   3.71      $\pm$   0.11      &   2.52      $\pm$   0.51      &   6.59      $\pm$   5.53      &   1.57      \\      
27         &  244.687016     &   -50.353573     &   5.46      $\pm$   0.73      &   2.87      $\pm$   1.67      &   -2.60     $\pm$   1.34      &   -8.20     $\pm$   1.54      &   -1.73     $\pm$   2.66      &   3.74      $\pm$   0.16      &   1.62      $\pm$   0.50      &   12.36     $\pm$   6.40      &   0.01      \\      
\hline 
\multicolumn{12}{c}{Extra central source} \\
\hline 
28$^{a,b}$ &  244.650894     &   -50.314579     &   6.17      $\pm$   0.35      &   8.48      $\pm$   1.17      &   -4.25     $\pm$   1.30      &   -9.31     $\pm$   1.32      &   -4.20     $\pm$   3.65      &   4.36      $\pm$   0.06      &   3.52      $\pm$   0.21      &   18.38     $\pm$   2.26      &   3.09      \\      
\hline
\end{tabular}

\begin{flushleft}
\begin{footnotesize}
~~~~~~~~~~~~~~$a$\,: Prominent sources associated with the central nebula; $b$\,: Did not have 8.0\,\micron\, photometry; 
$c$\,: Identified as \textquotedblleft G333.0494+00.0324B\textquotedblright\, (and also classified as a YSO) in 
\citet{lumsden13}. 
\end{footnotesize}
\end{flushleft}
\end{table}
\end{landscape}

\begin{table}
\centering 
\caption{Fitting results for compact \HII\, region}
\label{table_Radio} 
\begin{tabular}{@{}lcc}
\hline 
                        &  1280\,MHz & 843\,MHz \\
\hline 
2-D Gaussian fit        &            &          \\
size                    &  60.30\arcsec$\times$39.71\arcsec\,  &  69.41\arcsec$\times$76.17\arcsec\,  \\
Beam-deconvolved        &            &          \\
source size             &  23.35\arcsec$\times$18.31\arcsec\,  &  51.25\arcsec$\times$54.97\arcsec\,  \\
Position Angle (deg)    &  160.1     &  128.2   \\ 
Peak flux               &            &          \\
density (mJy beam$^{-1}$)  &  204$\pm$9  &  258$\pm$5    \\ 
Integrated flux         &            &          \\
density (mJy)           &  252$\pm$18 &  552$\pm$16   \\ 
\hline
\end{tabular}
\end{table}

\end{document}